\documentstyle[12pt,aaspp4]{article}
\begin{document}

\title{ROSAT X-ray Colors and Emission Mechanisms in Early-Type Galaxies}

\author{Jimmy A. Irwin and Craig L. Sarazin}
\affil{Department of Astronomy, University of Virginia, \\
P.O. Box 3818, Charlottesville, VA 22903-0818 \\
E-mail: jai7e@virginia.edu, cls7i@virginia.edu}

\begin{abstract}

The X-ray colors and X-ray--to--optical luminosity ratios $(L_X/L_B)$ of
61 early-type
galaxies observed with the {\it ROSAT} PSPC are determined.
The colors indicate that the X-ray spectral properties of galaxies vary
as a function of $L_X/L_B$.
The brightest X-ray galaxies have colors consistent with thermal emission
from hot gas with roughly the same metallicity of 50\% solar.
The spatial variation of the colors indicates that the gas temperature in
these galaxies increases radially.
Galaxies with medium $L_X/L_B$ also have spectral properties 
consistent with emission from hot gas. If a simple one-component thermal model 
is assumed to describe the $0.1-2.0$ keV X-ray emission in these galaxies,
then one possible explanation for the progressive decrease in $L_X/L_B$
among galaxies of this class could be the progressive decrease in metal
abundance of the X-ray emitting gas contained by the galaxies. However, stellar
X-ray emission may become a complicating factor for the fainter galaxies
in this medium $L_X/L_B$ class.

Galaxies with the lowest $L_X/L_B$ values appear to be lacking a hot
interstellar component. Their X-ray colors are consistent with those
derived from the bulges of the spiral galaxies M31 and NGC~1291. In M31
the X-ray emission is resolved into discrete sources, and is apparently
due primarily to low mass X-ray binaries (LMXBs).
We therefore suggest that the bulk of the X-ray emission in the faintest
ellipticals is also due to LMXBs. Previously, the X-ray spectra of X-ray 
faint galaxies had been found to be described by a hard component which
was attributed to LMXB emission, and a very soft component of unknown origin.
We show that the very soft component also likely results from LMXBs,
as a very soft component is seen in the X-ray spectra of the nearby LMXB
Her X-1 and LMXBs in the bulge of M31. If the X-ray emission in X-ray
faint galaxies is primarily from stellar sources, then a range in $L_X/L_B$
among these galaxies suggests that the stellar X-ray luminosity does
not scale with optical luminosity, at least for galaxies of low optical
luminosities. 
This could be the result of a decrease in the proportion of LMXBs
with decreasing optical luminosity, and/or the effects of fluctuations in
the small number of LMXBs expected.

\end{abstract}


\section{Introduction} \label{sec:intro}

One surprising discovery uncovered by the {\it Einstein} Observatory was
that elliptical and S0 galaxies are relatively bright, extended X-ray
sources (e.g., Forman et al.\ 1979; Forman, Jones, \& Tucker 1985;
Trinchieri \& Fabbiano 1985).
It soon was established that the source of these X-rays was thermal
emission from hot ($\sim$1 keV), diffuse interstellar gas, at least
for the brighter X-ray early-type galaxies.
However, several unsolved problems still remain concerning the X-ray
emission from elliptical and S0 galaxies. Although there is a strong
correlation between the X-ray and blue luminosities in these systems, the X-ray
luminosity can vary by as much as a factor of $50-100$ among galaxies with the
same blue luminosity (Canizares, Fabbiano, \& Trinchieri 1987;
Fabbiano, Kim, \& Trinchieri 1992).

Ellipticals and S0s with a low X-ray--to--blue luminosity ratio ($L_X/L_B$)
also have different spectral characteristics than their X-ray bright
(high $L_X/L_B$) counterparts.
 From {\it Einstein} observations,
Kim, Fabbiano, \& Trinchieri (1992) found that
the galaxies with the lowest $L_X/L_B$ exhibited an excess of very soft
X-ray emission.
{\it ROSAT} Position Sensitive Proportional Counter (PSPC) observations
of several
X-ray faint galaxies have confirmed that the X-ray emission in such galaxies
is not characterized by emission from hot, metal-enriched gas with a
temperature of $\sim$1 keV;
a two component model, in which one component has a
temperature around 0.2 keV and the other a temperature greater than
a few keV provides a better fit
(Fabbiano, Kim, \& Trinchieri 1994; Pellegrini 1994).
This result was confirmed by an {\it ASCA} observation of NGC~4382
(Kim et al.\ 1996).
The source of the soft $\sim$0.2 keV emission has been postulated to be
RS CVn binary systems, M star coronae, supersoft sources, or a low temperature
interstellar
medium (ISM), although none of these mechanisms seems to be entirely
responsible for the soft X-ray emission (Pellegrini \& Fabbiano 1994).
The hard component is thought to be emission from accreting compact
X-ray binaries;
this is the dominant X-ray component in most spiral galaxies
(Trinchieri \& Fabbiano 1985).

It is difficult to do detailed spectral modeling of these low $L_X/L_B$
systems because of their low X-ray count rates, even for the closest
and brightest examples.
Instead, in this paper we will analyze the broad band X-ray
``colors" of elliptical and S0 galaxies.
These ``colors'' are the ratios of the counts in three X-ray bands.
Whereas conventional spectral fitting
requires a minimum of about 1000 counts to constrain reasonably the
parameters of the model, only about 100 counts are needed to determine
the X-ray colors.
This allows much fainter X-ray galaxies to be analyzed and
compared to brighter galaxies, leading to a more comprehensive
investigation into the X-ray properties of early-type galaxies.
A similar technique of determining colors was used to study a large sample of
X-ray galaxies with {\it Einstein} data by Kim et al.\ (1992).
In this paper, we will determine the X-ray colors of early-type galaxies
observed with the {\it ROSAT} PSPC.
The increased sensitivity of {\it ROSAT} over {\it Einstein}
allows a larger sample to be analyzed and the X-ray colors to be determined
more accurately. In addition to integrated X-ray colors, we also
derive the X-ray colors as a function of galactocentric radius to explore
the variation of X-ray emission with position.
If the emission
mechanisms are different between X-ray faint and X-ray bright elliptical and
S0 galaxies, the integrated colors and color profiles should shed some light
on this difference.

This paper is organized as follows.
In \S~\ref{sec:galsamp}, we define our galaxy sample and give the relevant
properties of the galaxies, and in \S~\ref{sec:datared}, we discuss the data
analysis.
The X-ray luminosities and X-ray--to--optical luminosity ratios in hard and
soft bands are discussed in \S~\ref{sec:lxlb}.
The integrated X-ray colors of the galaxies are presented in
\S~\ref{sec:colors}, and are compared to simple models.
The radial profiles of the X-ray colors are given in
\S~\ref{sec:radcolor}.
In \S~\ref{sec:origin}, we discuss a number of different mechanisms for the
X-ray emission of elliptical galaxies of varying X-ray--to--optical
luminosities.
We show that the very soft component in X-ray faint galaxies is unlikely to
be due to stellar sources other than LMXBs or a warm ISM in
\S~\ref{sec:soft_alternatives}.
Finally, our conclusions are summarized in \S~\ref{sec:conclusions}.

\section{Galaxy Sample} \label{sec:galsamp}

In order to create an unbiased galaxy sample, we screened all the galaxies
in the {\it Third Revised Catalog of Galaxies} (de Vaucouleurs et al.\ 1991)
according to de Vaucouleurs T-type (T $\le -1$) and corrected blue apparent
magnitude
($m^0_B \le 12.5$). Using {\it HEASARC BROWSE}, we determined which galaxies
were observed with the {\it ROSAT} PSPC and available in the public archive.
Since both the sensitivity and resolution of the PSPC decrease appreciably
outside the rib support structure, we only include galaxies which are
contained within the inner edge of the rib ($\la 18^{\prime}$).
Furthermore, we excluded all galaxies with Galactic column densities above
$6 \times 10^{20}$ cm$^{-2}$, since soft X-rays are very heavily absorbed
at high hydrogen column densities. We excluded galaxies that are at the
center of rich clusters, since the X-ray emission from these galaxies is
contaminated by emission from hot intracluster gas known to be present in
such systems.
Finally, galaxies deeply buried
\clearpage
\begin{table}[ht]
\caption[Galaxy Sample]{}
\vskip8.00truein
\includegraphics{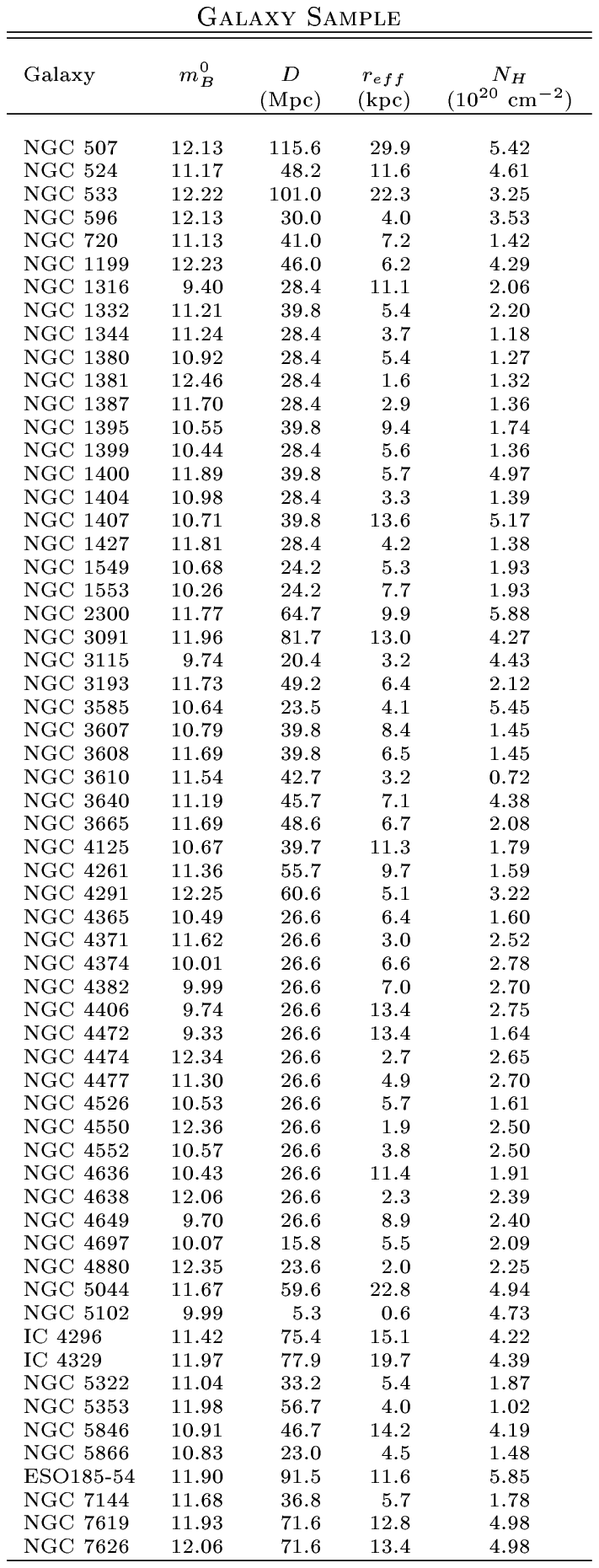}
\label{tab:galsamp}
\end{table}
\clearpage
\noindent  in the X-ray halo of another bright
galaxy, or unresolved from another X-ray source, or whose X-ray emission was
clearly
point-like, or which contained an active galactic nucleus according to
{\it Quasars and Active Galactic Nuclei}
(7th Edition; Veron-Cetty \& Veron~1996) were excluded.
These selection criteria yielded a total of 61 galaxies, whose properties
are listed in Table~\ref{tab:galsamp}.
The values for the blue total apparent magnitude $m_B^0$ were taken
from de Vaucouleurs et al.\ (1991).
The Galactic hydrogen column density to each galaxy is from
Stark et al.\ (1992).
The distances to the galaxies are also listed in Table~\ref{tab:galsamp}.
For most of the galaxies, distances were taken from
Faber et al.\ (1989);
the remaining distances were taken from Tully (1988).
The sole exception was ESO185-54, which did not appear in either of the above
sources. Its distance was taken from Sadler, Jenkins, \& Kotanyi (1989).
We have assumed a Hubble constant of $H_0 = 50$ km s$^{-1}$ Mpc$^{-1}$.

\section{Data Reduction} \label{sec:datared}

All the data were initially processed using Snowden's (1995) routines
in order to correct for particle and scattered solar X-ray background, and
for exposure and vignetting.
Periods of high background due to charged particles were removed by
filtering the data such that all time intervals with a Master Veto Rate
above 170 counts s$^{-1}$ were excluded.
For each data set, three images were created by combining pairs of
the seven energy band images produced by the Snowden (1995) routines
as follows:
R1L + R2 (approximately $0.11-0.41$ keV),
R4 + R5 (approximately $0.52-0.90$ keV), and
R6 + R7 (approximately $0.91-2.02$ keV).
Snowden's R3 band is located in the carbon absorption edge of the
detector, is poorly calibrated, and contains little independent
spectral information.
Nearly all the data were taken in the post-gain change mode
(1991 October 11).
However, any data
taken before the gain change were analyzed in the low gain bandpass (R1L)
for consistency.

For each of the three energy bands, counts were extracted from
concentric annular apertures $30^{\prime\prime}$ in width centered on the
peak of X-ray emission for each galaxy.
Unrelated X-ray sources within the field of view were identified by eye
and removed from the data.
In most cases, background counts
were obtained from an annular region $30^{\prime}-40^{\prime}$ in extent,
and were normalized to and subtracted from each source.
After subtracting background, the radial surface brightness profile of
the galaxies at large radii fluctuated around zero.
This provides evidence that the Snowden (1995) routines do a good
job of correcting for exposure, vignetting, and non-X-ray background.
In cases where the galaxy of interest was on the outskirts of
another galactic X-ray halo, a local background had to be carefully chosen.
Again, a surface brightness profile which fluctuated around zero at large
distances from the source confirmed the suitability of our background
region selection.

Next, we define our two X-ray colors as
\begin{equation} \label{eq:c21}
{\rm C21} =
\frac{\rm counts~in~0.52-0.90~keV~band}{\rm counts~in~0.11-0.41~keV~band}
\, ,
\end{equation}
and
\begin{equation} \label{eq:c32}
{\rm C32} =
\frac{\rm counts~in~0.91-2.02~keV~band}{\rm counts~in~0.52-0.90~keV~band}
\, .
\end{equation}
Although similar in spirit to
the X-ray colors used by Kim et al.\ (1992) for {\it Einstein} data,
the colors in these two studies are not directly comparable,
because of the different photon energy bandpasses and different responses
of {\it Einstein} and {\it ROSAT}.

For a given emission spectrum,
the observed C21 (and, to a lesser extent, C32) color depends
significantly on the amount of absorption due to Galactic hydrogen along
the line of sight to the galaxy.
We developed a technique to correct the colors for absorption,
based on the detailed spectral response of the {\it ROSAT} PSPC.
To calibrate the absorption corrections, we first determined the
emitted colors (C21 and C32) and the observed colors
(C21$^*$ and C32$^*$) for an extensive grid of
Raymond-Smith (1977) thermal models with varying abundances
(ranging from 20\% to 100\% of solar)
and temperatures
(ranging from 0.2 to 1.5 keV), and subject to
different values for the amount of foreground absorption.
The absorbing hydrogen columns considered ranged from
$N_H = 0.6$ to $6 \times 10^{20}$ cm$^{-2}$;
the upper limit corresponds to the upper limit allowed for inclusion
in our sample (\S~\ref{sec:galsamp}).
The spectral models were folded through the spectral response of
the PSPC using the XSPEC program.
We assumed that
the unabsorbed colors (C21, C32) could be related to the absorbed colors
(C21$^*$, C32$^*$) through relations of the form
\begin{equation} \label{eq:c21cor}
{\rm ln~C21} = {\rm ln~C21^*}
- A_1(N_H)~\left( \frac{N_H}{10^{21} \, {\rm cm}^{-2}} \right)
+ B_1(N_H)~\left( \frac{N_H}{10^{21} \, {\rm cm}^{-2}} \right)~{\rm ln~C21^*}
\end{equation}
\begin{equation} \label{eq:c32cor}
{\rm ln~C32} = {\rm ln~C32^*}
- A_2(N_H)~\left( \frac{N_H}{10^{21} \, {\rm cm}^{-2}} \right) \, ,
\end{equation}
where $A_1(N_H), B_1(N_H)$, and $A_2(N_H)$ are functions of the Galactic
column density $N_H$.
In equations (\ref{eq:c21cor}) and (\ref{eq:c32cor}), the terms with
the coefficients $A_1$ and $A_2$ would represent simple exponential
absorption if the spectral bands used were infinitely narrow.
The term with the coefficient $B_1$ includes the effect of the spectral
distribution of photons within the bands (as parameterized by the color)
on the effective optical depth.
For the harder C32 color, this additional correction was unnecessary.
The absorbed colors from our grid of spectral models and absorptions
were compared to those expected from
the correction equations (\ref{eq:c21cor}) and (\ref{eq:c32cor}).
A number of simple forms for the correction coefficient functions
$A_1$, $A_2$, and $B_1$ were tried.
Empirically, the following functions were found to work well.
Once the form of the functions was established, the best-fit
values were determined by least squares fitting to the emitted colors.
The resulting correction coefficient functions were
\begin{equation} \label{eq:A1}
A_1 (N_H) = 4.57 \left( \frac{N_H}{10^{21} \, {\rm cm}^{-2}} \right)^{-0.30}
\end{equation}
\begin{equation} \label{eq:B1}
B_1(N_H) = \log \left[ 3.85
\left( \frac{N_H}{10^{21} \, {\rm cm}^{-2}} \right)
- 0.21 \right] \, ,
\end{equation}
\begin{equation} \label{eq:A2}
A_2 = 0.21 \, .
\end{equation}
This empirically-determined relation was quite effective in correcting
the colors for absorption, producing only a 6\% rms error in C21 and a 2\%
rms error in C32 when comparing the absorption-corrected colors to the
original unabsorbed colors derived from the spectral models.
The rms errors became substantially larger when models
with hydrogen column densities greater than $6 \times 10^{20}$ cm$^{-2}$
were included in the calculation of the correction coefficient functions.
It is for this reason that we have excluded all galaxies with column densities
above $6 \times 10^{20}$ cm$^{-2}$ from our sample.
For the remainder of this paper, we will refer to the
absorption-corrected colors as C21 and C32.

One of the aims of this paper is to study the differing X-ray emission
mechanisms in galaxies with differing X-ray luminosities.
X-ray luminosities from the {\it Einstein} Observatory are
available for many of our galaxies.
However, for consistency with the spectral properties derived from
the {\it ROSAT} PSPC data, we also derived X-ray luminosities from
the same data.
The X-ray luminosity $L_X$ of each galaxy was determined for band 1
($0.11-0.41$ keV; the ``soft'' s band), and for bands 2 and 3 combined
($0.52-2.02$ keV; the ``hard'' h band).
The count rates for these bands were converted to physical fluxes and
luminosities assuming that the emission spectrum was a Raymond-Smith
thermal model with a temperature of 0.8 keV and an abundance of 0.5 solar,
using separate model normalizations for hard and soft bands.
The fluxes are not strongly dependent on the emission spectrum
assumed.
The fluxes were corrected for intervening absorption, based on the
Galactic hydrogen column toward the galaxy (Table~\ref{tab:galsamp}).
For consistency, the luminosities were all determined for a
circular aperture whose radius was five optical effective radii.
The effective radius of each galaxy $r_{eff}$, which contains one-half
of the projected optical luminosity of the galaxy, was taken from
de Vaucouleurs et al.\ (1991)
and is listed in Table~\ref{tab:galsamp}.
Unrelated X-ray sources which fell within
the extraction region were removed.

The X-ray luminosities for both bands are listed in Table~\ref{tab:xprop}.
Some of the galaxies were not detected in one or both of the bands.
For these, Table~\ref{tab:xprop} lists the
$1\sigma$ confidence upper limit on the X-ray luminosity, preceded by
a $<$.
We also determined the blue optical luminosities $L_B$ of the galaxies from
their magnitudes and distances in Table~\ref{tab:galsamp}.
The
\begin{table}[tbp]
\caption[X-Ray Properties]{}
\vskip8.25truein
\includegraphics{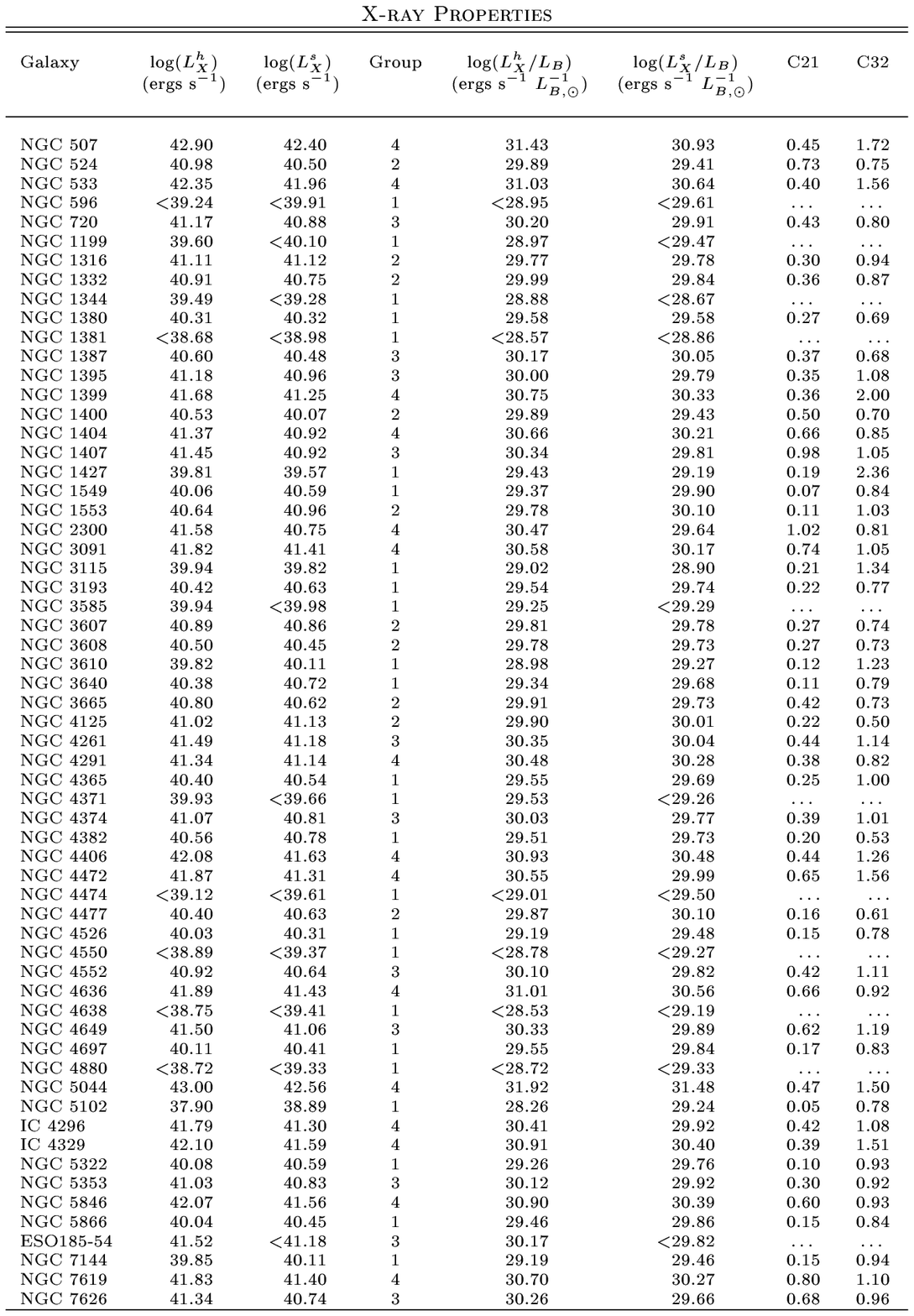}
\label{tab:xprop}
\end{table}
\clearpage
\begin{table}[tbp]
\caption[Average Group Properties]{}
\vskip1.5truein
\includegraphics{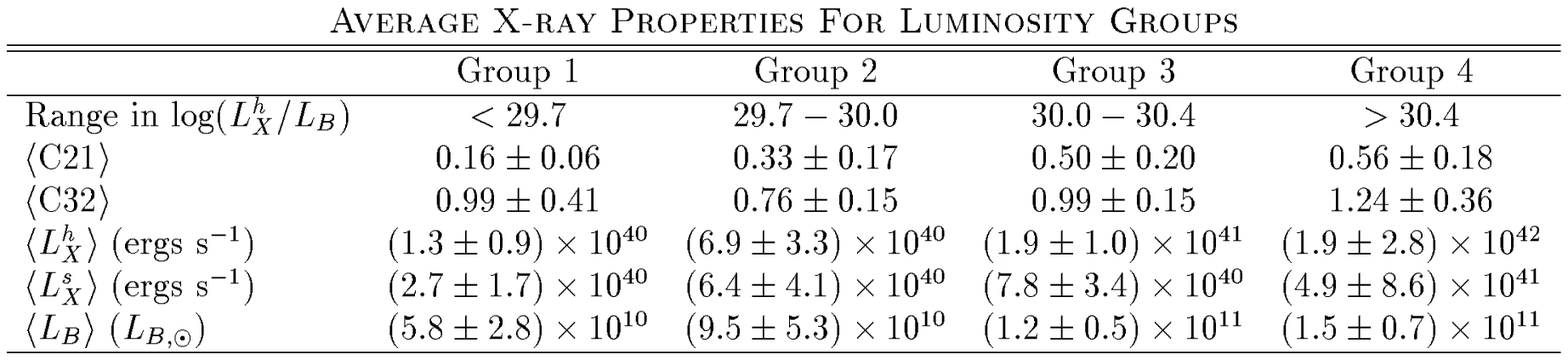}
\label{tab:group_prop}
\end{table}
\noindent  X-ray--to--optical luminosity ratios $L_X/L_B$
are also listed in Table~\ref{tab:xprop} for each of the two X-ray bands.
The galaxies were grouped into four luminosity classes by their
$0.52-2.02$ keV X-ray--to--blue luminosity ratio ($L^h_X/L_B$).
We have followed the convention of Kim et al.\ (1992) of labeling the
X-ray faintest galaxies as Group 1 and the X-ray brightest galaxies
as Group 4, although the boundaries of the four groups differ from theirs.
Group 4 galaxies have $\log(L^h_X/L_B) \ge 30.4$, Group 3 galaxies have
$30.0 \le \log(L^h_X/L_B) < 30.4$, Group 2 galaxies have
$29.7 \le \log(L^h_X/L_B) < 30.0$, and Group 1 galaxies have
$\log(L^h_X/L_B) < 29.7$.
These boundaries were chosen so that each group contained roughly the
same number of galaxies, and as will be shown below, the same general
spectral characteristics.
The average properties of the galaxies in our sample within each of
these groups are listed in Table~\ref{tab:group_prop}.
For each property, the number given is the mean value and the standard
deviation within the group (rather than the error of the mean).
The standard deviation is useful as a representation of the scatter
in the property among the members of the group.

\section{X-ray/Optical Luminosity Relation for Soft and Hard X-Ray Bands}
\label{sec:lxlb}

\begin{figure}[htbp]
\vskip6.30truein
\hskip0.3truein
\includegraphics{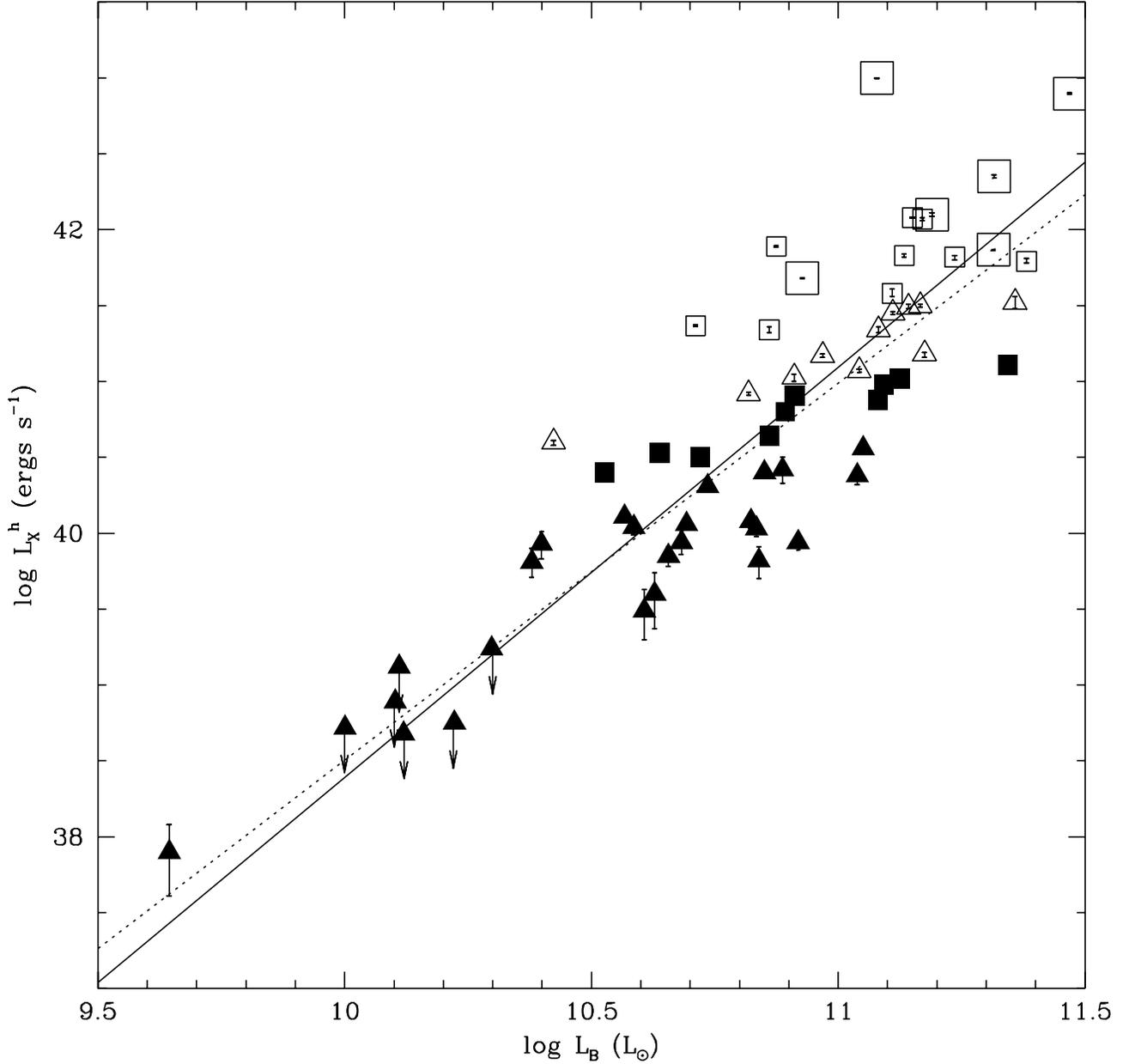}
\caption[Hard X-rays]{
The logarithm of the hard band ($0.52-2.02$ keV) X-ray luminosity $L_X^h$ is
plotted versus the logarithm of the blue optical luminosity $L_B$
for all of the galaxies in the sample.
Group 1, 2, 3, and 4 galaxies are denoted by filled triangles, filled
squares, open triangles, and open squares, respectively.
Symbols with one-sided vertical arrows are upper limits.
The error bars or upper limits on the X-ray luminosity are $1\sigma$.
The six Group 4 galaxies located at the centers of groups
(Table~\protect\ref{tab:groups})
are shown as large open squares.
The best-fit power-law relation is
$L_X^h \propto L_B^{2.70 \pm 0.20}$
if all galaxies are included (solid line),
and $L_X^h \propto L_B^{2.48 \pm 0.19}$ if the six Group 4 galaxies located in
the centers of groups are excluded (dotted line). \label{fig:lxhlb}}
\end{figure}

The hard band X-ray luminosities or upper limits from Table~\ref{tab:xprop}
are plotted as a function of the optical luminosity in Figure~\ref{fig:lxhlb}.
Using all the galaxies in the sample, we found that
the unweighted best-fit relation between the hard band X-ray luminosity
($L_X^h$) and the optical luminosity was
$L_X^h \propto L_B^{2.70 \pm 0.20}$.
Galaxies with only upper limits on their X-ray luminosity were included
in the fit, using
the parametric EM algorithm within the survival analysis package {\it ASURV}
(Isobe, Feigelson, \& Nelson 1986).
This is not an unbiased determination of the X-ray
luminosities of optically selected ellipticals, because of the unknown
selection function associated with {\it ROSAT} PSPC observations of the
galaxies.
Still, our relation is consistent with
the White \& Davis (1997) result, once they excluded dwarf ellipticals and
M32.
M32 appears as a point source in the {\it ROSAT} High Resolution
Imager,
and most likely contains a small number of X-ray binaries or
a micro-AGN at its center (Eskridge, White, \& Davis 1996).
The dispersion in our $L_X^h$ vs.\ $L_B$ relation was 0.55 dex.

Several of the most luminous X-ray galaxies sit at the centers
of groups of galaxies, and are likely to have their X-ray
emission enhanced by group gas.
Six of the galaxies in our sample located at the centers
of groups have integrated X-ray colors
distinctly different from the rest of the Group 4 galaxies, as will be shown
below (\S~\ref{sec:colors}, Figure~\ref{fig:c21c32}).
These six galaxies are listed in Table~\ref{tab:groups}, which also gives
some of their X-ray properties.
\begin{table}[tbp]
\caption[Group Centers]{}
\begin{center}
\begin{tabular}{lccccc}
\multicolumn{6}{c}{\sc Galaxies at the Centers of Groups} \cr
\hline \hline
Galaxy & Group Affiliation & C21 & C32 & log($L_X^h/L_B$) & log($L_X^s/L_B$)\\
&&&& (ergs s$^{-1} L_{B,\odot}^{-1}$) & (ergs s$^{-1} L_{B,\odot}^{-1}$) \\
\hline
NGC~507 & NGC~507 Group & 0.45 & 1.72 & 31.43 & 30.93 \\
NGC~533 & GH14 & 0.40 & 1.56 & 31.03 & 30.64 \\
NGC~1399 & Fornax Cluster & 0.36 & 2.00 & 30.75 & 30.33 \\
NGC~4472 & Virgo South & 0.65 & 1.56 & 30.55 & 29.99 \\
NGC~5044 & NGC~5044 Group & 0.47 & 1.50 & 31.92 & 31.48 \\
IC~4329 & IC~4329 Group & 0.39 & 1.51 & 30.91 & 30.40 \\
\hline
\end{tabular}
\end{center}
\label{tab:groups}
\end{table}
They are also indicated by open squares in Figure~\ref{fig:lxhlb}.
If we remove these six galaxies,
we obtain $L_X^h \propto L_B^{2.48 \pm 0.19}$,
with a dispersion of 0.49 dex.
Although this relation is flatter than the relation derived using all
the galaxies, it is still somewhat steeper than the value derived by
White \& Davis (1997), once they removed the cooling flow galaxies
NGC~4486, NGC~4696, and NGC~1399 from their sample.
The best-fit relations with and without the 6 group center
galaxies are shown in Figure~\ref{fig:lxhlb}.
It should be noted that there are other galaxies in our sample that lie
at the center of groups (e.g., NGC~2300, NGC~3607, NGC~5846), but these
groups are generally poor, and the integrated X-ray colors appear normal.

\begin{figure}[htbp]
\vskip6.30truein
\hskip0.3truein
\includegraphics{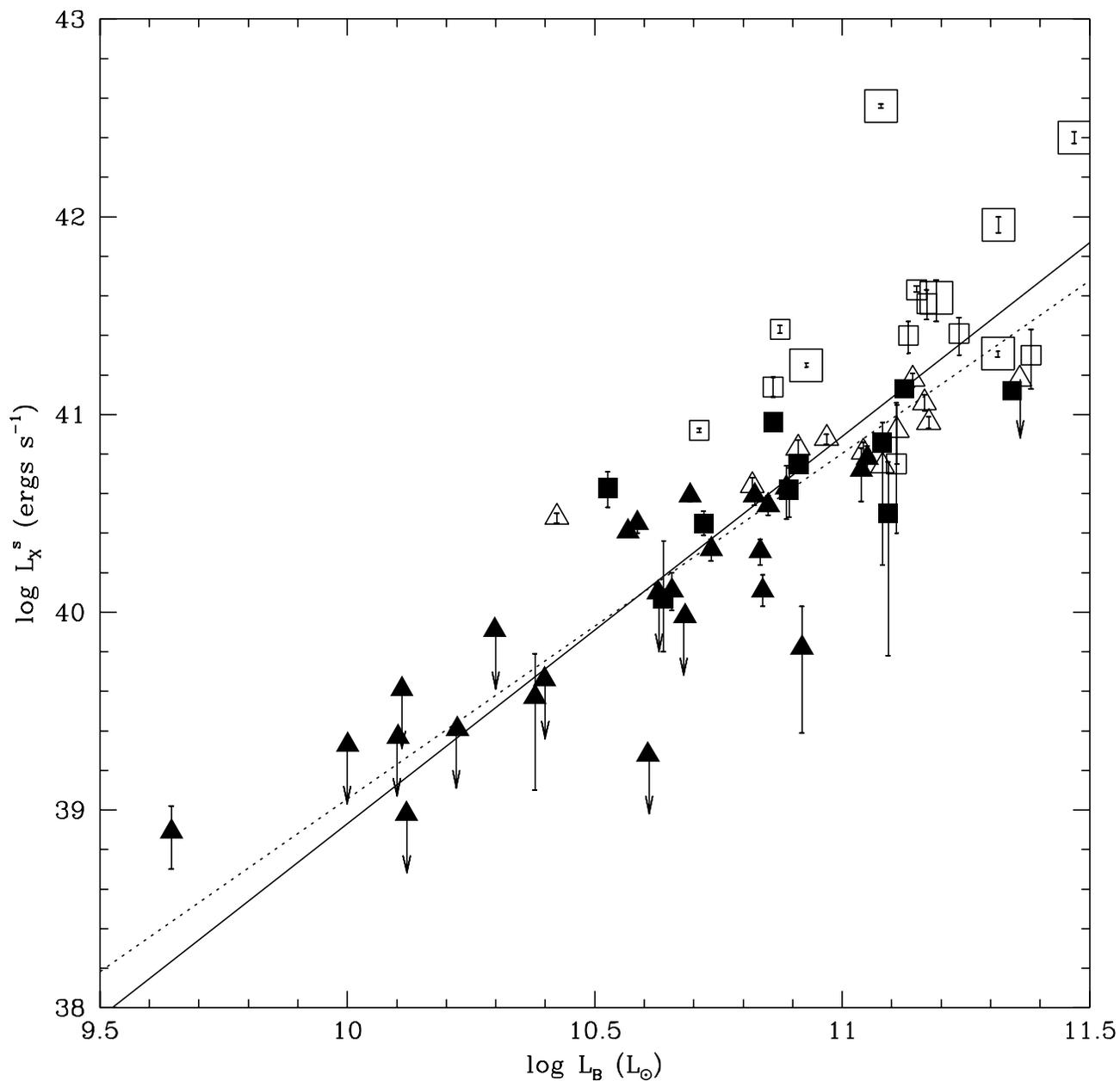}
\caption[Soft X-rays]{
The logarithm of the soft band ($0.11-0.41$ keV) X-ray luminosity $L_X^s$ is
plotted versus the logarithm of the blue optical luminosity $L_B$
for all of the galaxies in the sample.
The notation is identical to that in
Figure~\protect\ref{fig:lxhlb}.
The best-fit power-law relations are $L_X^s \propto L_B^{1.94 \pm 0.17}$
(solid line, whole sample) and $L_X^s \propto L_B^{1.75 \pm 0.16}$
(dotted line, group galaxies excluded). \label{fig:lxslb}}
\end{figure}

In Figure~\ref{fig:lxslb}, the soft band X-ray luminosity $L_X^s$ is plotted
versus the blue optical luminosity $L_B$.
The unweighted best-fit
relation is $L_X^s \propto L_B^{1.94 \pm 0.17}$ with a dispersion
of 0.45 dex.
After excluding the six group center galaxies discussed above, we obtained a
best-fit relation of
$L_X^s \propto L_B^{1.75 \pm 0.16}$
with a dispersion of 0.39 dex.
In the soft band, the X-ray---optical correlation is much
flatter than for the hard band.
The best-fit relations with and without the group center
galaxies are shown in Figure~\ref{fig:lxslb}.

\begin{figure}[htbp]
\vskip6.30truein
\hskip0.3truein
\includegraphics{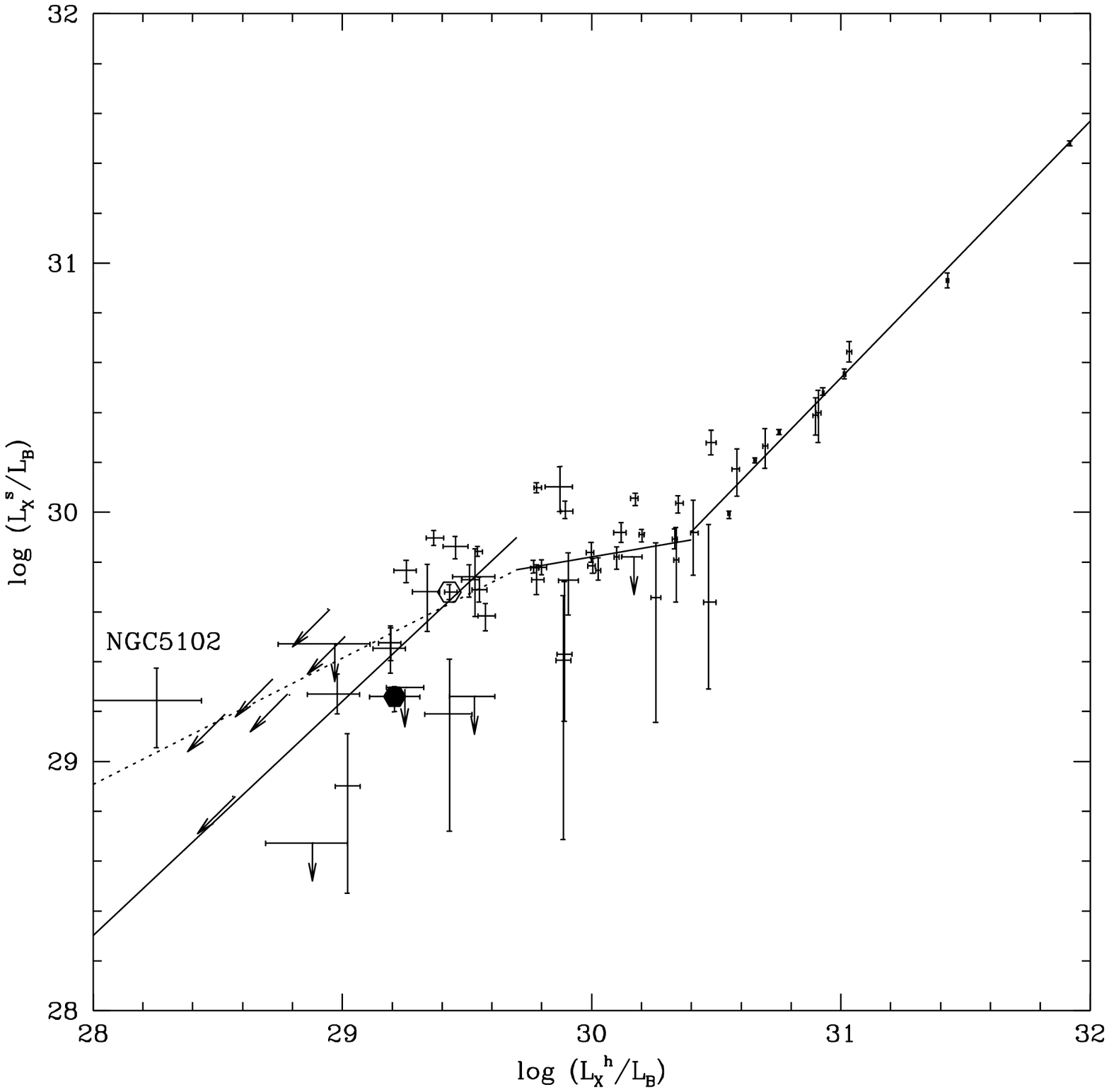}
\caption[Hard vs. Soft]{
The logarithm of the soft band X-ray--to--optical ratio
$\log(L_X^s/L_B)$ is plotted against the logarithm of the
hard band ratio $\log(L_X^h/L_B)$ for all of the 61 galaxies in our survey.
The solid lines show the best-fit linear relations for
Group 1, Group 2 and 3, and Group 4 galaxies.
In the case of Group 1 galaxies, NGC~5102 was excluded from
this fit.
The dotted line is the best-fit
relation for Group 1 galaxies if NGC~5102 is included.
Symbols with one-sided vertical arrows were detected in the hard band only, and
symbols with one-sided slanted arrows were not detected in either band.
The bulges of M31 and NGC~1291 are shown as a filled and open hexagon,
respectively. \label{fig:hard_soft}}
\end{figure}

Figure~\ref{fig:hard_soft} shows the soft band X-ray---optical ratio
log($L_X^s/L_B$) versus the hard band ratio log($L_X^h/L_B$).
There is a tight, linear relation between
$\log(L_X^s/L_B$) and $\log(L_X^h/L_B$) for galaxies with
$\log(L_X^h/L_B) > 30.4$ (Group 4).
However, for fainter X-ray galaxies in the range
$29.7 < \log(L_X^h/L_B) < 30.4$, the slope flattens
significantly;
these galaxies all have $\log(L_X^s/L_B) \approx 29.8$.
(The few galaxies in this region of
the plot which exhibit a lower value of $L_X^s/L_B$ all have Galactic column
densities which are near the upper limit for
our sample, so it is possible that their soft X-ray emission is underestimated
because of absorption.
At any rate, the
upper error bars on these galaxies are consistent with the $L_X^s/L_B$
values for the rest of the Group 2 and 3 galaxies.)
Finally, the X-ray faintest galaxies with
$\log(L_X^h/L_B) < 29.7$ (Group 1) again show a crudely linear relation between
$L_X^s/L_B$ and $L_X^h/L_B$.

To quantify these trends, we derived the best-fit slope between
$\log(L_X^s/L_B)$ and $\log(L_X^h/L_B)$ for the Group 4,
Groups 2 and 3, and Group 1 galaxies for which emission was detected in both
bands.
Galaxies with upper limits were excluded in the fit, although all the galaxies
were consistent with the trends found above.
For Group 4 galaxies, we find
$L_X^s/L_B \propto (L_X^h/L_B)^{1.03 \pm 0.09}$.
For Group 2 and 3 galaxies
we find $L_X^s/L_B \propto (L_X^h/L_B)^{0.18 \pm 0.22}$.
Finally, for Group 1 galaxies we find
$L_X^s/L_B \propto (L_X^h/L_B)^{0.50 \pm 0.19}$.
This relation becomes nearly linear if NGC~5102 is excluded,
$L_X^s/L_B \propto (L_X^h/L_B)^{0.94 \pm 0.32}$.
This steeper relation for the Group 1 galaxies is more consistent with the
upper limits for galaxies which were undetected.
Inspection of the $0.11-0.41$ keV image of NGC~5102 shows a considerable
number of foreground/background sources near the position of NGC~5102,
so it is not clear if the source at the position of NGC~5102 is actually
emission from the galaxy or just a positional coincidence with an unrelated
source.
The best-fit relations for Group 1 (both with and without NGC~5102), for Groups
2 and 3, and for Group 4 are shown in Figure~\ref{fig:hard_soft}.

\section{Integrated X-ray Colors} \label{sec:colors}

In order to compare the average spectral properties of the galaxies in our
sample, the integrated colors representing emission from the whole
galaxy were determined for each galaxy in which X-ray emission was
detected in all three bands. The size of the extraction region was chosen
to be that radius which provided the highest signal-to-noise ratio in C21.
Although this meant discarding all counts in bands 2 and 3 outside of this
radius (which sometimes amounted to a large fraction of the total emission in
the two harder bands), it
was necessary to ensure that the colors C21 and C32 represented the emission
from the same physical space within the galaxy. In addition, the extraction
radius was always set to at least 1$^{\prime}$ to reduce the effects of the
energy-dependent Point Spread Function (PSF) of the PSPC on the colors.
After the raw colors were determined for each galaxy, they were corrected for
Galactic absorption according to the procedure outlined in
\S~\ref{sec:datared}.

\begin{figure}[htbp]
\vskip6.30truein
\hskip0.3truein
\includegraphics{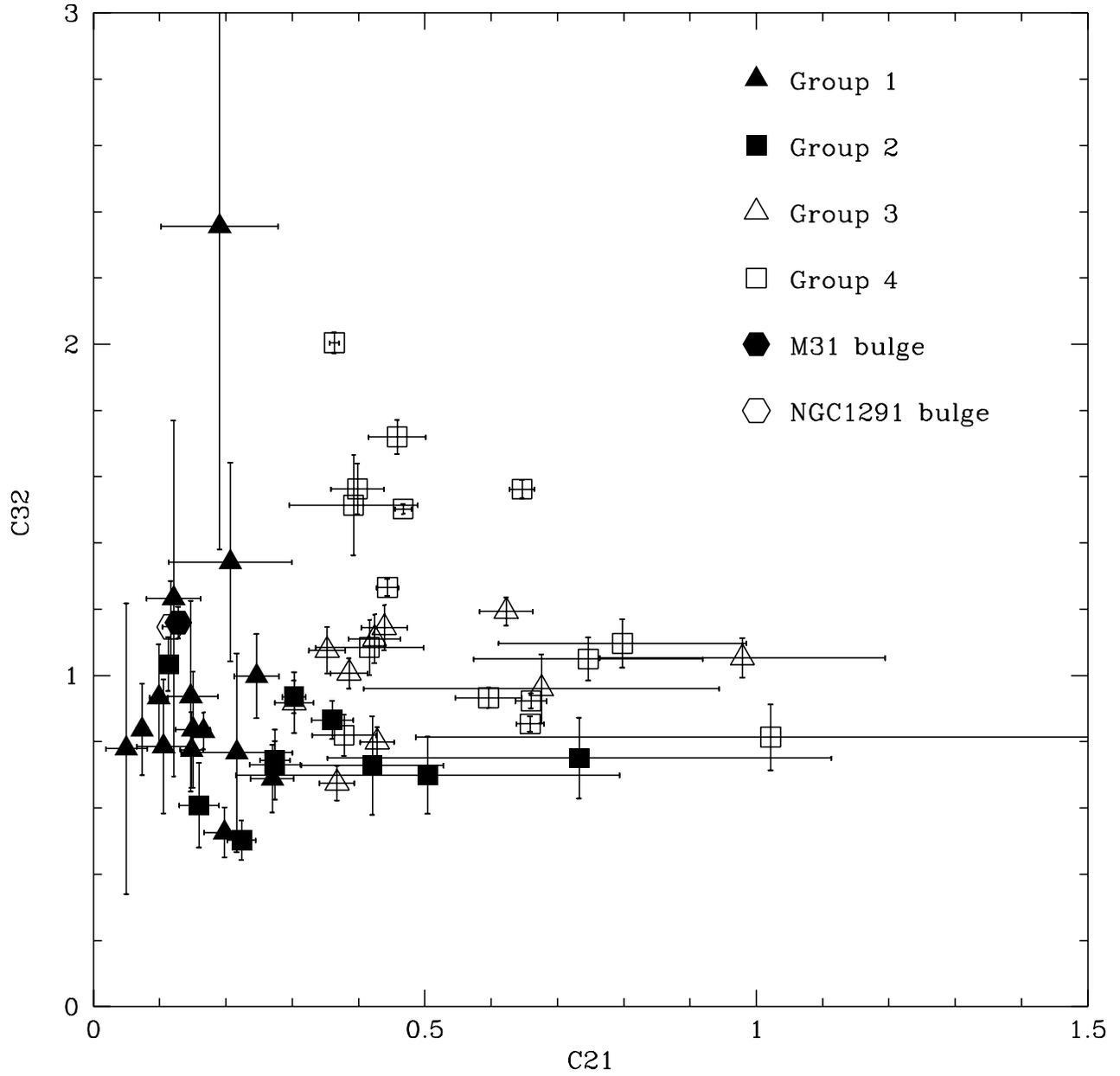}
\caption[X-ray Color Plot]{
A plot of the integrated C21 and C32 colors for the 50 galaxies which had
emission in all three bands, with the $1\sigma$ errors. The filled hexagon
represents the bulge of M31 and the open hexagon represents the bulge
of NGC~1291. \label{fig:c21c32}}
\end{figure}

A plot of C21 vs.\ C32 is shown in Figure~\ref{fig:c21c32} for all galaxies in
the survey with detectable emission in all bands, along with the $1\sigma$
error bars.
The values for the integrated colors are also given in the
last two columns of Table~\ref{tab:xprop}.
It is obvious from Figure~\ref{fig:c21c32} that the X-ray faintest Group 1
galaxies (filled triangle) are separated from
the X-ray brightest Group 4 (open squares).
All Group 1 galaxies with the exception
of NGC~1427 at (C21, C32$)=(0.19, 2.36)$ have C21 colors less than 0.3 and
C32 colors consistent with $0.5-1.0$. In contrast, the Group 4 galaxies
have C21 colors greater than 0.3 and C32 colors above 0.8. Group 2 (filled
squares) and Group 3 (open triangles) galaxies provide for a smooth
transition in colors between the Group 1 and 4 galaxies.
The fact that X-ray
faint galaxies have relatively more very soft X-ray emission (lower C21)
than X-ray bright galaxies was hinted at by Kim et al.'s~(1992) {\it Einstein}
study, although their statistics were too poor to prove this for each
galaxy individually.
Pointed observations by {\it ROSAT} showed this to be true for a few of the
faint X-ray galaxies with the highest fluxes
(e.g., Fabbiano, Kim, \& Trinchieri 1994; Pellegrini 1994).
This study shows that all X-ray faint early-type galaxies with
detectable emission have excess very soft emission.

\begin{figure}[htbp]
\vskip6.30truein
\hskip0.3truein
\includegraphics{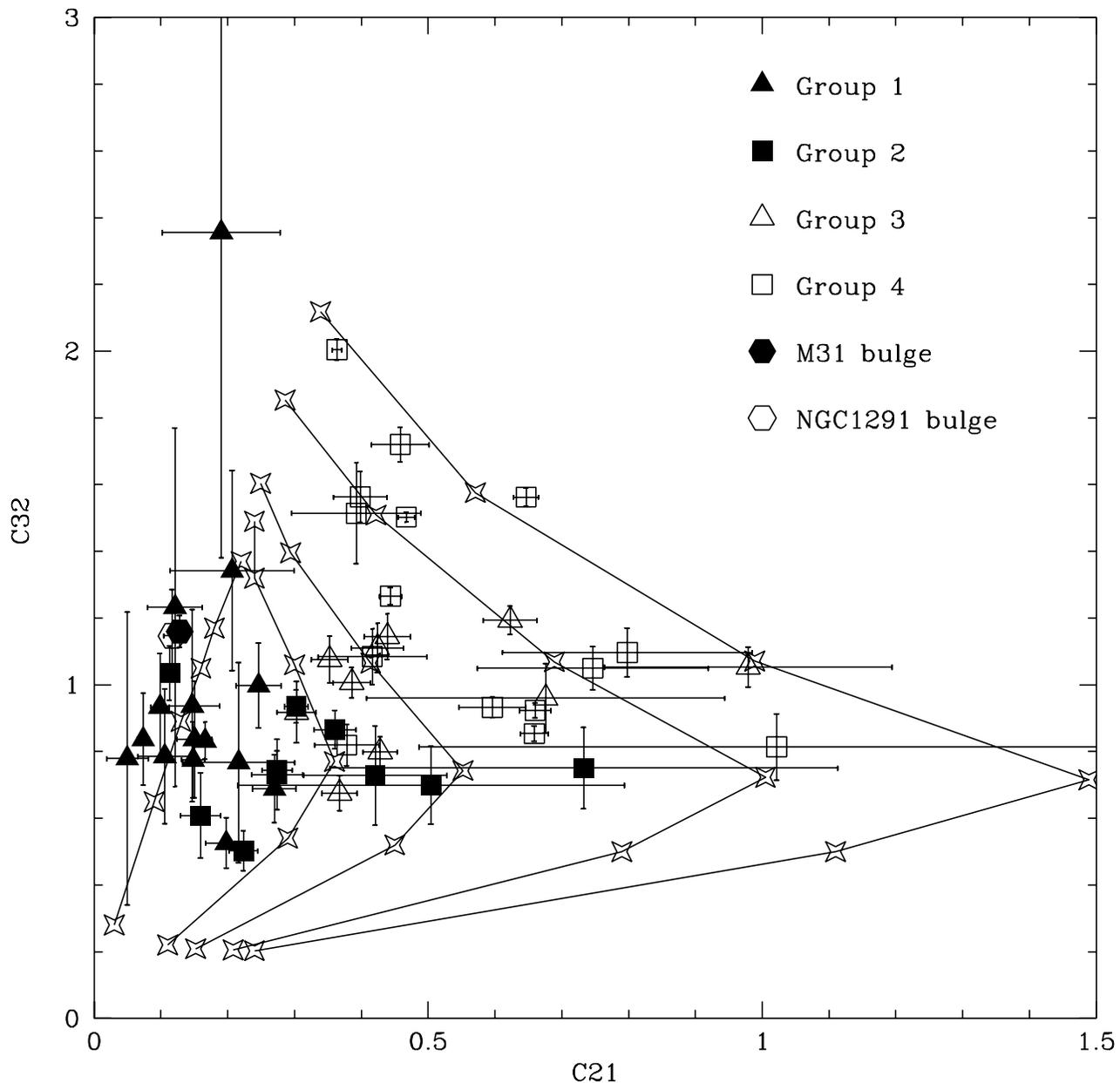}
\caption[X-ray Color Plot with Models]{
The C21 and C32 colors from Figure~\protect\ref{fig:c21c32} are shown with
lines indicating the loci for simple thermal models
for the emission.
The solid lines represent the colors predicted from a variety
of Raymond-Smith models, with each track representing a different
metallicity: zero abundance (left), 10\%, 20\%, 50\%, and 100\% (right) of
solar. Along a given track, each symbol represents a different temperature:
0.2 (bottom), 0.4, 0.6, 0.8, 1.0, 1.5 (top) keV. \label{fig:c21c32em}}
\end{figure}

It is useful to compare the colors derived from the data to the colors
predicted from various single component thermal models once the models
have been folded through the spectral response of the PSPC.
Since the true X-ray emission mechanism of early-type galaxies over the
entire X-ray spectrum is more complicated than a simple one component model,
these comparison models are intended only as a general description of
the emission in the $0.1-2$ keV energy range. At higher energies, stellar
emission quite possibly contributes significantly to the X-ray emission.
We present the colors for single component thermal models because
more complex spectral models with more than two free parameters
are difficult to represent in a two-dimensional plot with data points.
The colors expected from Raymond-Smith thermal models are shown
as solid lines in Figure~\ref{fig:c21c32em}, which shows the
same integrated color data for the galaxies as in Figure~\ref{fig:c21c32}.
Each track represents a different metallicity, and each point along a given
track represents a different temperature.
The X-ray bright Group 3 and Group 4 galaxies lie in the same
region of C21-C32 space as thermal models with temperatures
between 0.6 and 1.5 keV and abundances between 20\% and
100\% of solar.
This agrees with previous analyses of the X-ray spectra of
the brighter galaxies
(e.g., Buote \& Canizares 1994, and the references in the Introduction).
It is worth mentioning that abundance and temperature gradients may
affect the conclusions drawn from single component spectral models.
The integrated colors represent the average emission properties of the
galaxies, and the effect of gradients will be discussed in
\S~\ref{sec:radcolor}.

It is possible that some of the galaxies in the sample harbor AGN despite
the fact that they are not listed in the Veron-Cetty \& Veron (1996)
catalog. However, if AGN emission were important in these galaxies,
their X-ray colors would most likely be significantly different than
galaxies without AGN. The lack of conspicuous outliers on the color-color
diagram suggests that it is unlikely that any galaxies with strong AGN
X-ray emission remain in the sample.

A subgroup of the Group 4 galaxies
have lower C21 and higher C32 colors than
the rest of the Group 4 galaxies, indicative of higher temperatures as shown
by the model tracks.
As noted in \S~\ref{sec:lxlb},
these higher temperature Group 4 galaxies all sit at
the center of a group of galaxies.
These galaxies are identified in Table~\ref{tab:groups}.
It seems likely that a deeper potential well due to the group is responsible
for the higher temperatures and different colors than the rest of
the Group 4 galaxies.

Most of the X-ray faintest Group 1 galaxies have colors which are
consistent with very low abundance ($< 20\%$ solar)
thermal models with a variety of temperatures.
A few of the Group 1 galaxies have colors inconsistent with
even zero metallicity and single temperature optically thin
thermal emission.
The plausibility of low metallicity gas being the
source of the X-ray emission in these X-ray faint galaxies will be discussed
below (\S~\ref{sec:origin_group1}).

\begin{figure}[htbp]
\vskip6.30truein
\hskip0.3truein
\includegraphics{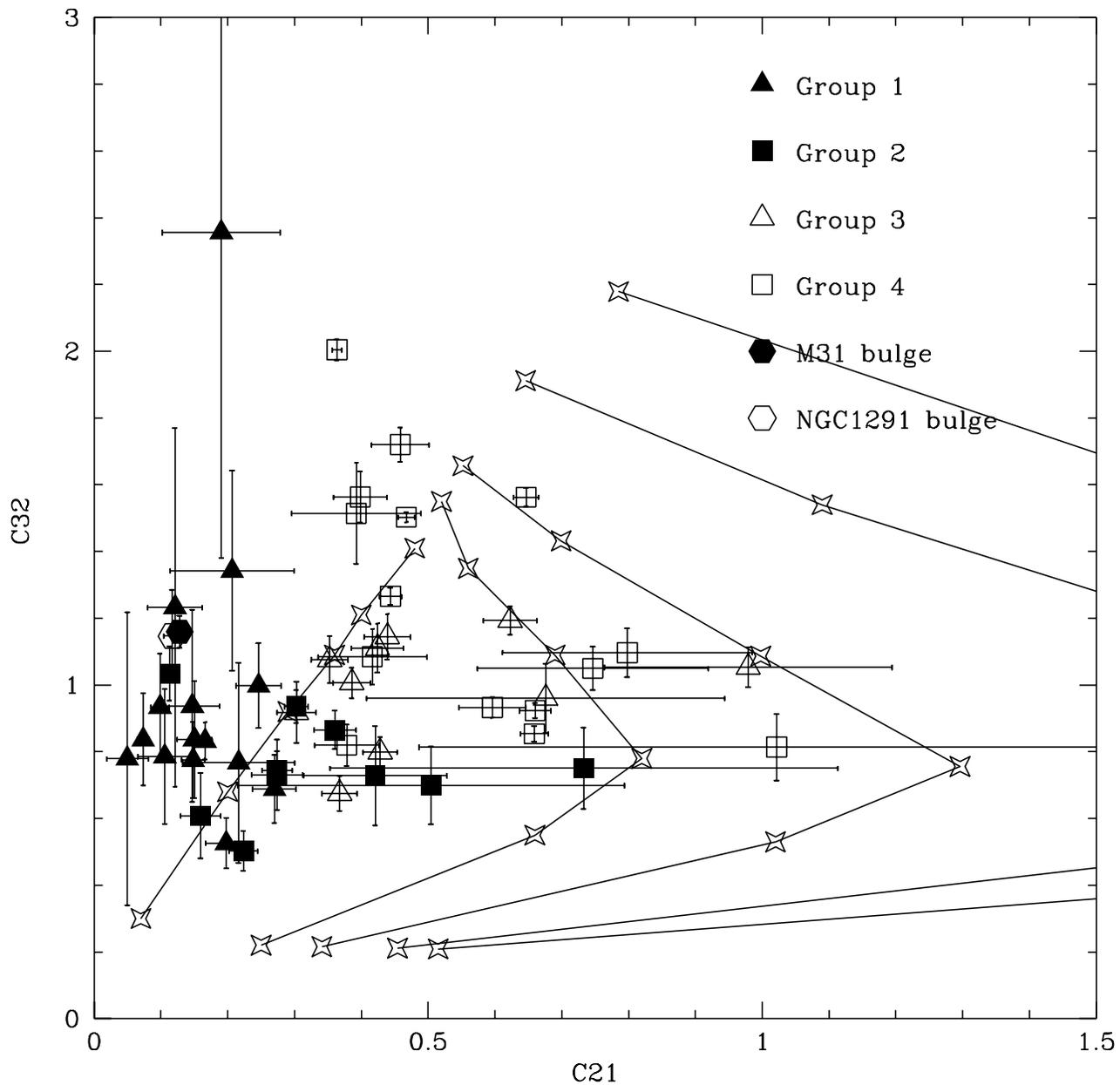}
\caption[X-ray Color Plot with Absorption]{
Same as Figure~\ref{fig:c21c32}, except the model thermal spectra
have an extra foreground absorbing column density of
$10^{20}$ cm$^{-2}$.
The addition of the extra absorption
shifts C21 to larger values (to the right),
but hardly affects C32. \label{fig:c21c32abs}}
\end{figure}

Davis \& White (1996) find that absorption due to material which
is intrinsic to the galaxy is not substantial in their sample of 30 galaxies.
We have examined this possibility for our sample.
We computed the colors of the same thermal models shown in
Figure~\ref{fig:c21c32em}, but added a foreground absorbing
screen with a column density of $10^{20}$ cm$^{-2}$.
The absorbed model colors are shown as solid lines in
Figure~\ref{fig:c21c32abs}.
By comparing to Figure~\ref{fig:c21c32em}, we see that
C32 is hardly changed by the addition
of the absorption component, but C21 is shifted significantly to the right.
None of the galaxies in our sample fall within the region of C21-C32
space occupied by the absorbed models with abundances greater than 20\% of
solar.
Therefore, we conclude that foreground column densities
intrinsic to the galaxies $\ga 10^{20}$ cm$^{-2}$ are not
present, unless the abundance in these galaxies is very low, and then only in
the brightest X-ray galaxies.
This comparison also serves as an indication that our procedure
for correcting the observed colors for absorption is reasonably
accurate.

\section{Radial Color Variation} \label{sec:radcolor}

In addition to the integrated colors, we also calculate the colors as a
function of radius, C21($r$) and C32($r$), for each galaxy with enough counts
to do so.
In order to extract the maximum amount of spatial information
consistent with the PSF of the PSPC and with reasonable
statistical accuracy, we followed the following procedure.
Counts were accumulated in annuli with widths of 30$^{\prime\prime}$, starting
with an innermost circle with
this radius.
Succeeding annuli were combined until the colors from the
combined ring could be determined with a signal-to-noise
ratio of at least 4.
When this occurred, we stopped accumulating counts in that
ring, and started another.
Again, we combined 30$^{\prime\prime}$ annuli until this next ring had a
signal-to-noise ratio of at least 4, and so on.
This was repeated out to the last radial bin in which the
signal-to-noise ratio was greater than two.
This was done to avoid creating radial bins which might be dominated
by systematic errors in the background subtraction or vignetting correction,
rather than the statistical errors. 
Note that the last bin does not necessarily satisfy the
25\% accuracy criterion for the colors.
We required that each ring be at least 1$^{\prime}$ in radius or
width in order to reduce the effects of the energy dependent PSF of the PSPC.

\begin{figure}[htbp]
\vskip6.40truein
\hskip0.3truein
\includegraphics{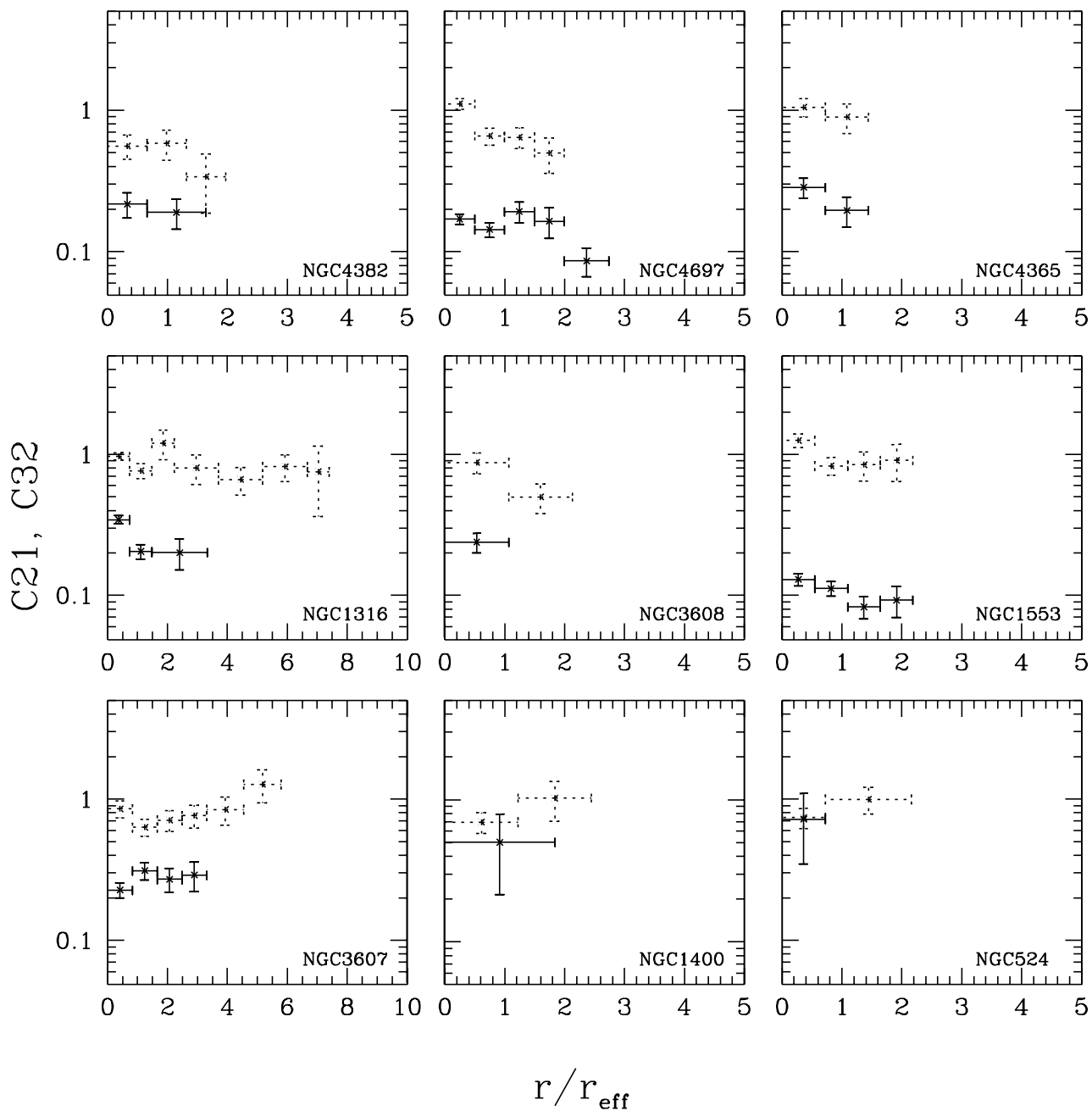}
\caption[Color Profiles]{
The radial profiles of the C21 (solid lines) and C32 (dotted lines) X-ray
colors for galaxies with at least two spatial C32 bins.
The colors are plotted versus the projected radius in
units of the optical effective radius.
The galaxies are ordered by increasing $L_X^h/L_B$. \label{fig:profiles}}
\end{figure}

The radial color profiles are presented in Figure~\ref{fig:profiles} for all
galaxies for which at least two spatial bins were obtained for C32.
As was the case with the integrated colors, the radial color profiles also are
distinctly different between different groups.
\begin{figure}[htbp]
\vskip6.60truein
\hskip0.3truein
\includegraphics{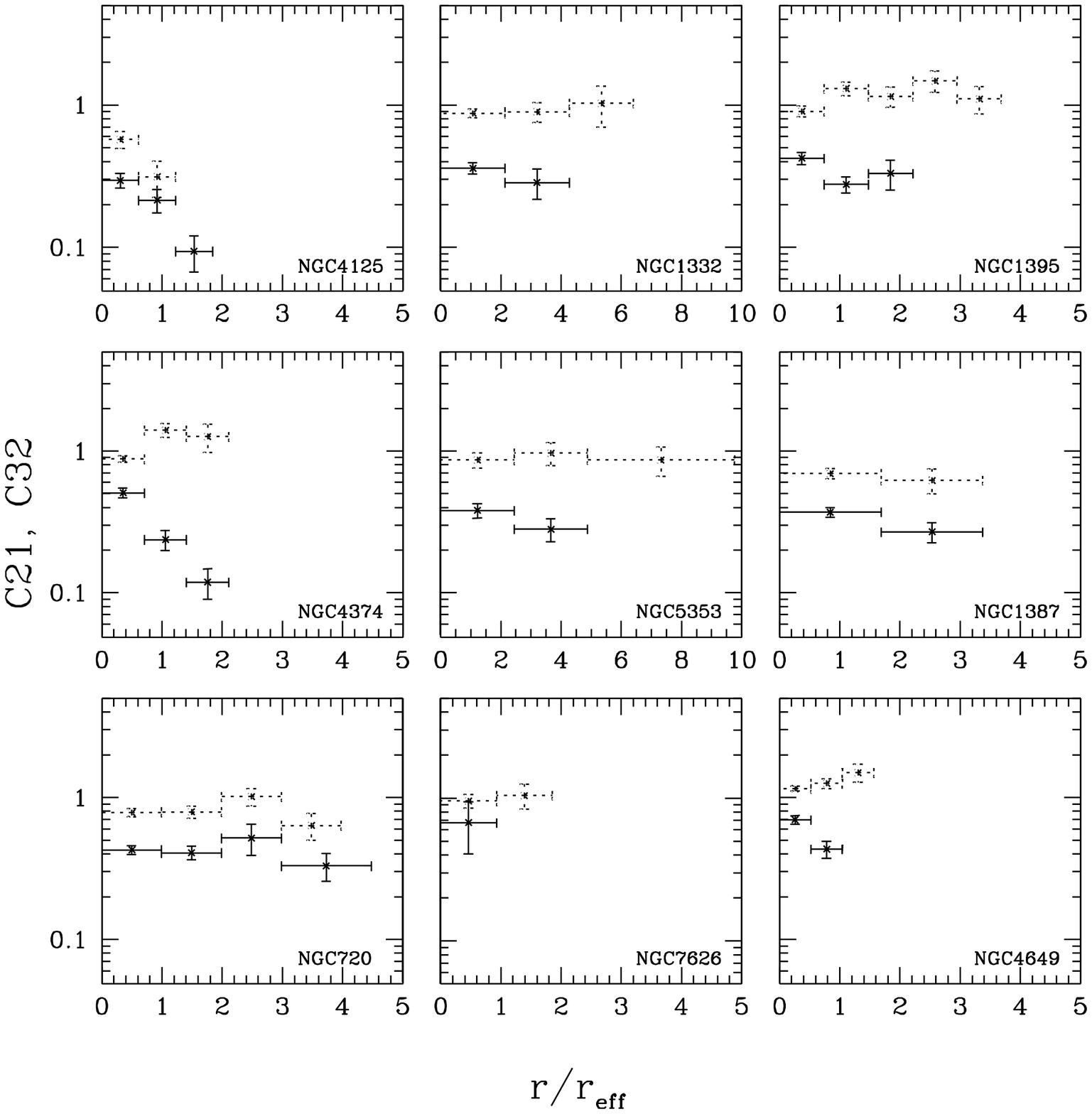}
\begin{center}
Fig. 7. --- Continued
\end{center}
\end{figure}
\begin{figure}[htbp]
\vskip6.60truein
\hskip0.3truein
\includegraphics{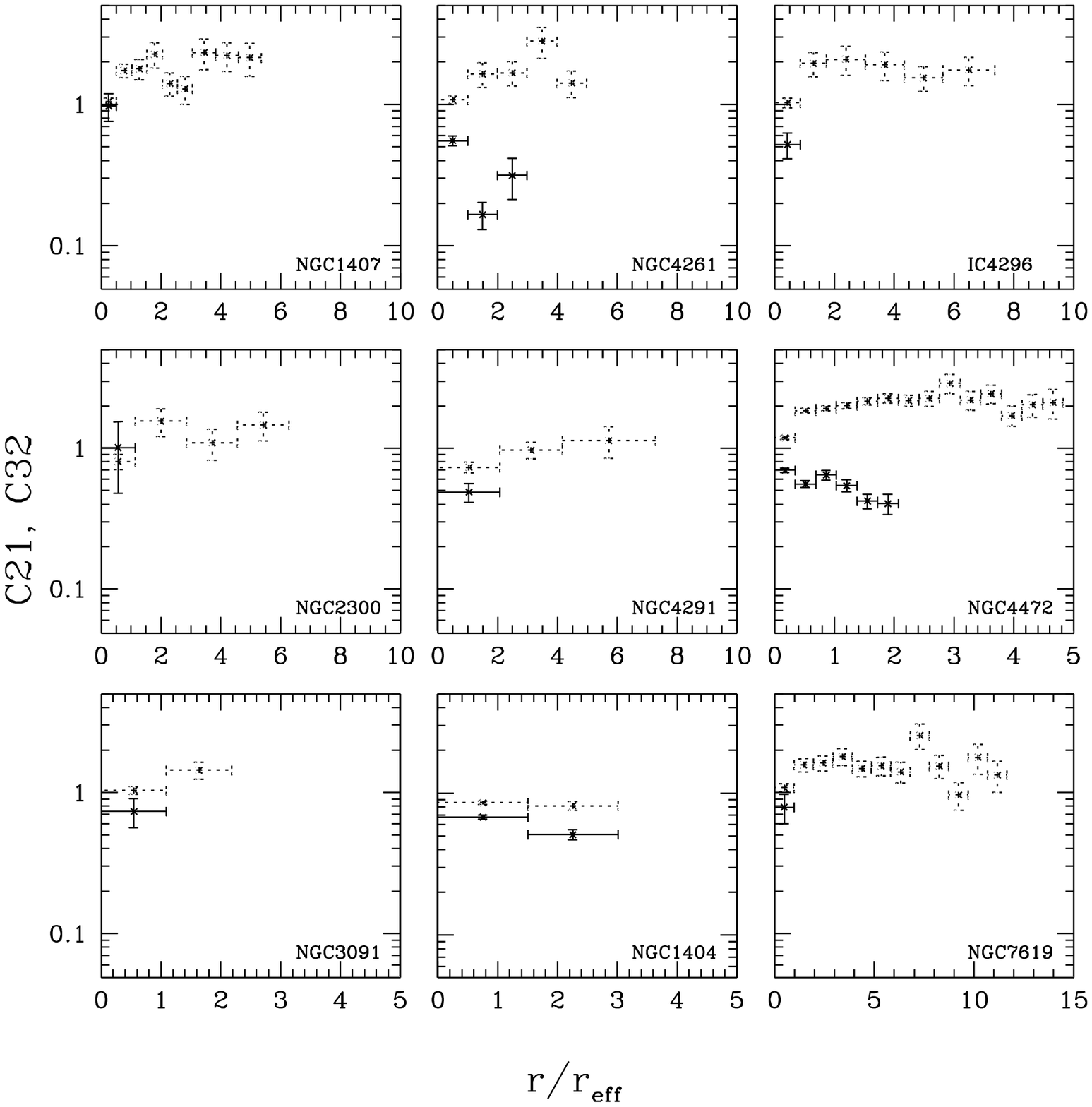}
\begin{center}
Fig. 7. --- Continued
\end{center}
\end{figure}
\begin{figure}[htbp]
\vskip6.60truein
\hskip0.3truein
\includegraphics{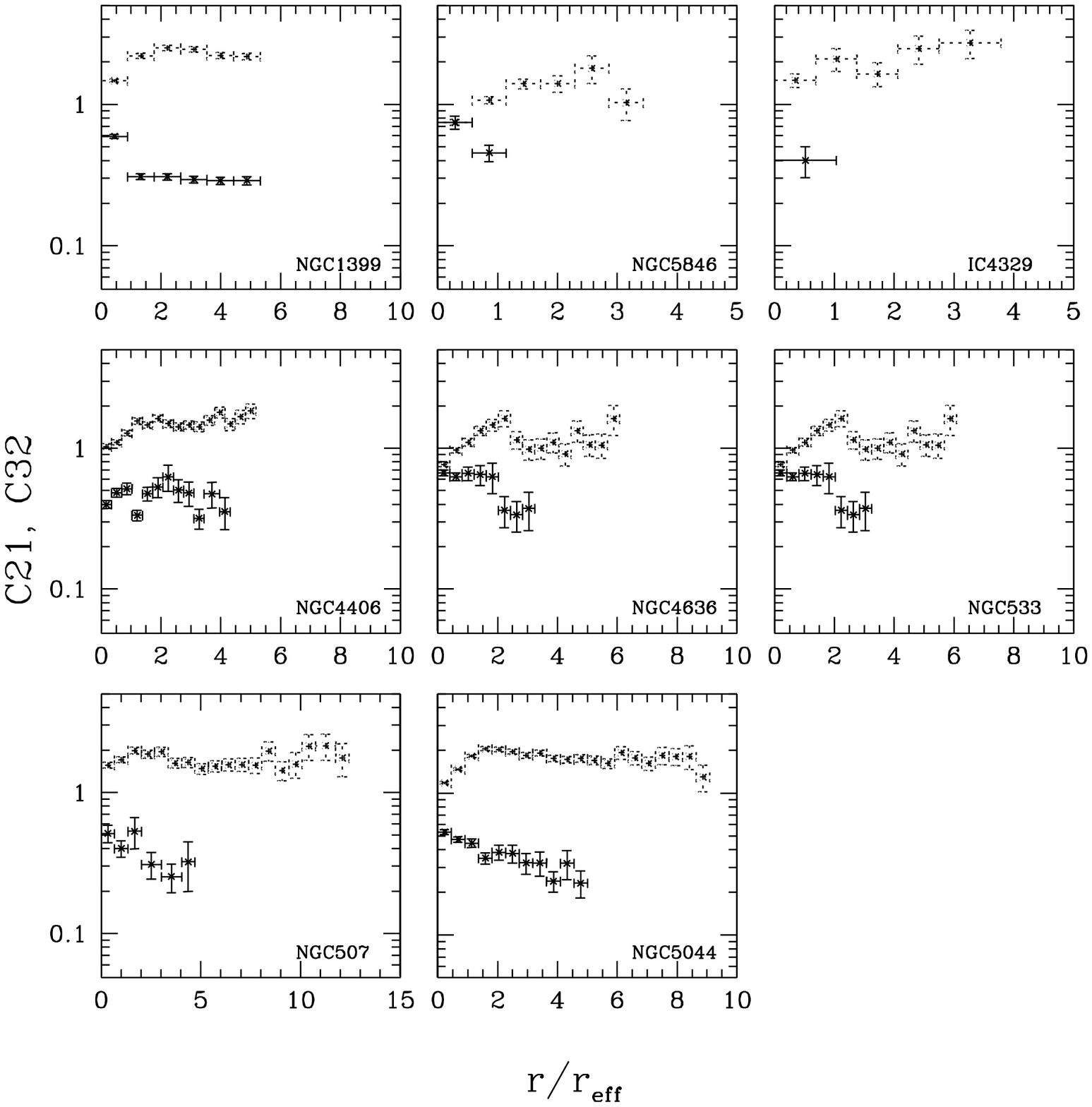}
\begin{center}
Fig. 7. --- Continued
\end{center}
\end{figure}
Although all the galaxies show a general decreasing radial trend in C21, the
behavior of the C32 radial profiles is a function of the X-ray--to--optical
luminosities.
The X-ray brightest Group 4 galaxies in general show an
increase in C32 out to a few effective radii, before leveling off or decreasing
somewhat at large radii.
Group 3 galaxies show the same trend, although some galaxies in this group have
a flat radial C32 profile.
Group 2 galaxies show a flat or somewhat decreasing
radial C32 profile.
Finally, the few X-ray faintest Group 1 galaxies for which a profile could
be determined had a constant or somewhat decreasing C32 profile and a roughly
constant C21 profile. 

\begin{figure}[htbp]
\vskip6.30truein
\hskip0.3truein
\includegraphics{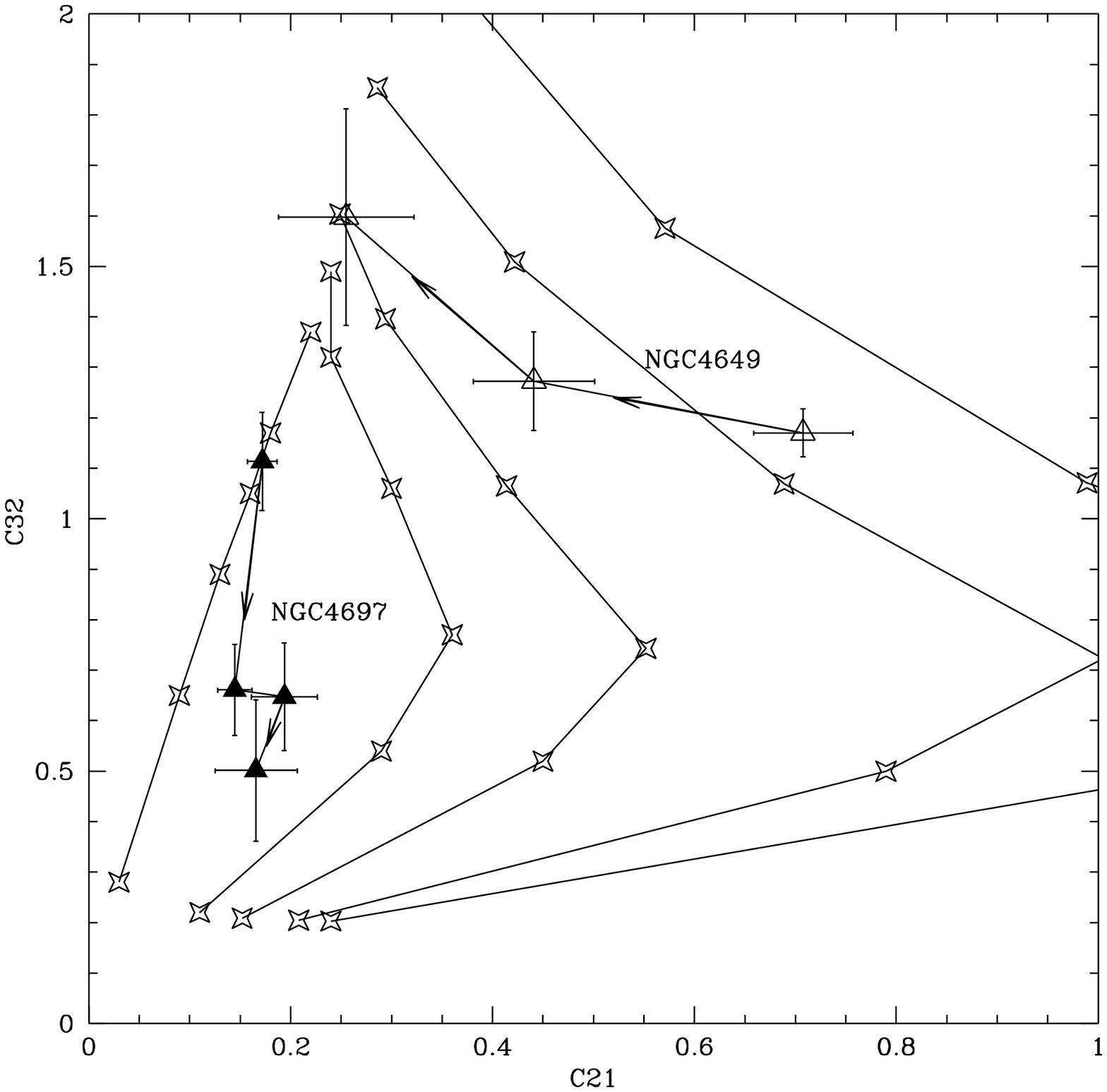}
\caption[C21-C32 Profiles]{
The spatial variation of the X-ray colors of two
example galaxies are shown in the C21-C32 plane.
The data points give the colors, while the arrows connecting
them show the direction of increasing radius.
The galaxies are labeled.
NGC~4649 is an X-ray bright, Group 3 galaxy, while
NGC~4697 is an X-ray faint Group 1 galaxy.
The model tracks are the same as
in Figure~\protect\ref{fig:c21c32em}. \label{fig:c21c32profile}}
\end{figure}

Figure~\ref{fig:c21c32profile} shows the radial color profiles of
two selected galaxies on the C21-C32 diagram.
The colors from a given galaxy are connected by solid lines, with arrows
indicating the direction of increasing radius.
We have chosen NGC~4649 as a suitable object to
represent the higher X-ray luminosity galaxies of
Groups 3 and 4.
As the galactocentric distance increases, the color change is
nearly consistent with a track of constant metallicity and
increasing gas temperature.
Such a radial temperature variation has been found previously
from detailed spectral fits to {\it ROSAT} PSPC data
on X-ray bright galaxies such as
NGC~4636 (Trinchieri et al.\ 1994),
NGC~4472 (Forman et al.\ 1993), and
NGC~507 (Kim \& Fabbiano 1995).
This drop in temperature at the center of the galaxy can be interpreted as
evidence for a cooling flow in these galaxies.
To the extent that the color tracks differ from those for
a constant abundance, they suggest that in general the iron abundance in
the gas decreases with radius.
Again, a similar trend has been found from detailed {\it ASCA}
X-ray spectra for the more luminous X-ray galaxies
(Mushotzky et al.\ 1994). 
Other galaxies in these groups show color variations which are
consistent with an initial increase in temperature with increasing radius
at small radii;
at larger radii, the X-ray colors remain constant within the
errors.
These color profiles suggest that cooling is very significant in
the central regions, but that the gas is isothermal in the outer
regions.

The situation is quite different for the few Group 1 which have enough
counts to determine the colors as a function of radius,
as shown by the colors profile of NGC~4697
in Figure~\ref{fig:c21c32profile}.
In this galaxy, the X-ray
colors do not vary in the way expected for increasing gas
temperatures with increasing radii.
On the contrary, if the color variation were interpreted in
terms of a single temperature thermal model, one would require
that the gas temperature decrease and iron abundance increase
with radius.
Moreover, many of the colors are inconsistent with any single thermal model,
including an iron abundance of zero.
The problems raised by the colors and color variations of Group 1 galaxies
are discussed further in \S~\ref{sec:origin_group1}.

\section{The Origin of the X-ray Emission in Early-Type Galaxies}
\label{sec:origin}

The C21-C32 plot (Figure~\ref{fig:c21c32}) and the $\log(L_X^s/L_B$) versus
$\log(L_X^h/L_B$) relation
(Figure~\ref{fig:hard_soft})
both suggest that the X-ray emission in elliptical and S0 galaxies is complex,
and not the result of one mechanism.
Here, we attempt to understand the diversity in X-ray spectral properties among
galaxies with varying X-ray--to--optical luminosities.

\subsection{Group 4 Galaxies} \label{sec:origin_group4}

Given their high X-ray luminosities and the fact that the nearest giant
elliptical galaxies are all at the approximate distance of the Virgo cluster,
Group 4 includes the galaxies with the
highest observed X-ray fluxes.
As a result, these galaxies have been studied in the most detail in the past.
In particular, the brightest Group 4 galaxies have high signal-to-noise spectra
from {\it ROSAT} or {\it ASCA} which can be
analyzed in detail.
Not surprisingly, our integrated colors and color profiles are consistent with
the results of these spectral studies.
The X-ray emission of the Group 4 galaxies is apparently
predominantly thermal X-ray emission from hot gas with temperatures of
$0.6-1.5$ keV and metallicities
$\sim$50\% of solar.
The tight linear correlation found between $\log(L_X^s/L_B$) and
$\log(L_X^h/L_B$) (Figure~\ref{fig:hard_soft}) also suggests
that a single emission mechanism, thermal emission by diffuse
gas, is predominant in these galaxies.

In a survey of early-type galaxies observed with {\it ASCA},
Matsumoto et al.\ (1997) found a hard ($5-10$ keV), presumably stellar
component in addition to a softer ISM component in nearly all the
galaxies observed. However, the $0.5-4.5$ keV luminosity of this
hard component was an order of magnitude less than that of the soft
ISM component in the X-ray brightest galaxies. After adopting an average
hard--to--soft component normalization from the Matsumoto et al.\ (1997)
sample, this hard component was added to the models. The addition of
the hard component to the existing models changed the model colors by no
more than 15\% for spectral models with temperatures 0.6 keV and greater.
For models with temperatures below 0.6 keV, the colors were affected somewhat
more, but not to the level where they became consistent with the observed
colors.

In general, the color profiles indicate that the temperature of the gas
increases with increasing galactocentric radius, suggesting that a cooling flow
is present in most of these high $L_X/L_B$ systems.

The tight linear correlation found between $\log(L_X^s/L_B$) and
$\log(L_X^h/L_B$) (Figure~\ref{fig:hard_soft}) and the nonlinear increase of
the X-ray luminosity in both bands with the optical
luminosity implies that any stellar X-ray emission is completely swamped by the
gaseous emission in the {\it ROSAT} band.

Many of the most luminous X-ray galaxies in Group 4 are located
at the centers of groups (Table~\ref{tab:groups}).
The higher X-ray luminosities of these galaxies may reflect
gas which was trapped in the gravitational potential of the
groups, with the brighter X-ray galaxies sitting in the deeper potential wells.
Similarly, the harder X-ray colors of these galaxies may reflect
the deeper potential well of the group.

\subsection{Group 2 and Group 3 Galaxies} \label{sec:origin_group23}

Below $\log(L_X^h/L_B$) = 30.4, the $\log(L_X^s/L_B$)
vs.\ $\log(L_X^h/L_B$) relation flattens considerably
(Figure~\ref{fig:hard_soft}).
One explanation of this leveling off of $L_X^s/L_B$ would be
that the emission in the soft X-ray band in these galaxies
is stellar in nature, and thus directly proportional to the optical
luminosity.
However, this would not explain why $L_X^s/L_B$ continues to decrease again
once log($L_X^h/L_B$) drops below 29.7.
In addition, analyses of {\it ROSAT} and {\it ASCA} X-ray
spectra of a few of these galaxies
(e.g., Davis \& White 1996; Matsumoto et al.\ 1997)
show that the bulk of the emission is thermal emission from hot gas and
subsolar abundances, much like Group 4 galaxies.
Unless the integrated stellar emission from these galaxies mimics the spectrum
from hot gas,
it is unlikely that we are seeing the stellar X-ray emission which must
underlie the X-ray emission in all early-type galaxies at some level.
Furthermore, the $L_X^h/L_B$ value expected from a stellar component,
derived from the Matsumoto et al.\ (1997) sample (see
\S~\ref{sec:origin_group1} below), is 29.4, an order of magnitude less than
where the flattening begins. Although stellar emission may become significant
in the faintest Group 2 galaxies ($L_X^h/L_B \sim 29.7$), it is unlikely
that stellar emission is important for the brighter Group 2 and 3 galaxies.

\begin{figure}[htbp]
\vskip6.30truein
\hskip0.3truein
\includegraphics{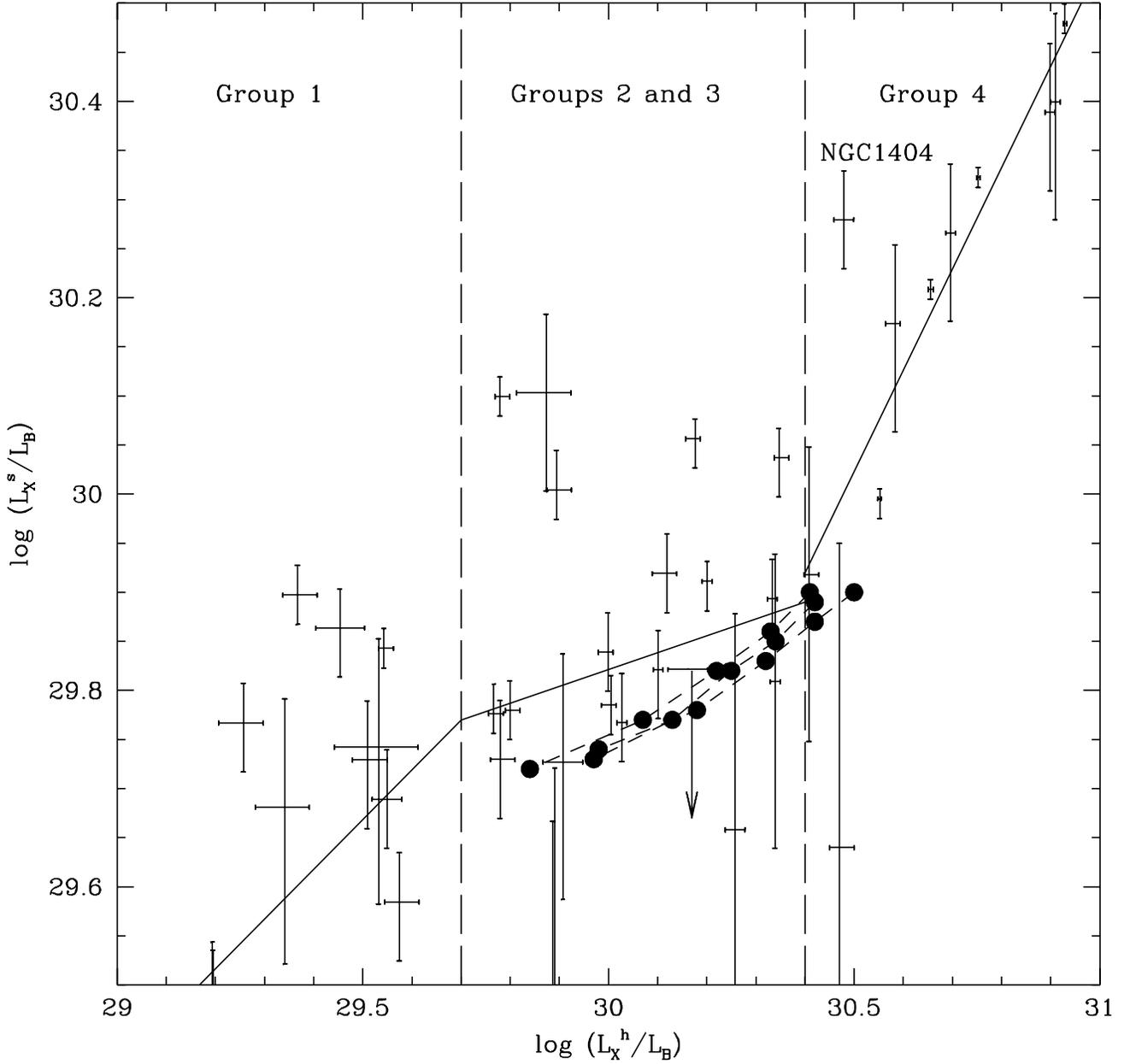}
\caption[Abundances]{
An enlarged view of the center of Figure~\protect\ref{fig:hard_soft}.
The data point corresponding to NGC~1404 is indicated.
The three dashed lines show the predicted $\log(L_X^s/L_B$) and
$\log(L_X^h/L_B$) values for Raymond-Smith models with
temperatures of 0.8 (top), 0.4, and 0.6 (bottom) keV, and abundances which
vary along each curve from 10\% of solar (left) to 50\% of solar (right).
The models assume a constant ratio of the volume emission measure of the
gas to the optical luminosity. \label{fig:abund}}
\end{figure}

The C21-C32 diagram (Figure~\ref{fig:c21c32}) provides an alternate explanation
for this flattening.
As $L_X^h/L_B$ decreases, C21 decreases while C32 remains roughly constant.
As the thermal emission models of Figure~\ref{fig:c21c32em} show,
this effect could be the result of variations in the abundance.
To test this, we have computed the luminosity expected in both the soft and hard
X-ray bands for a series of one component Raymond-Smith models with a constant
emission measure but a variety of temperatures and metallicities.
The X-ray luminosities were divided by a constant arbitrary blue luminosity
which yielded
$\log(L_X^s/L_B$) and $\log(L_X^h/L_B$) values which best matched the average
values of the Group 2 and 3 galaxies.
The results are shown in Figure~\ref{fig:abund} for
models with temperature of 0.4, 0.6, and 0.8 keV and abundances
which ranged from 10\% to 50\% of solar.
The three temperature tracks are shown as dashed lines.

We find that the hard-to-soft X-ray luminosity ratio
varies little with temperature between 0.4 and 0.8 keV, with differences in
this ratio being driven by abundance differences.
As the abundance decreases, the hard X-ray luminosity decreases significantly
faster than the soft X-ray luminosity for a given emission measure.
The reason for this is apparently that the decrease in abundance reduces the
strength of the iron $L$-line complex around 1 keV
which dominates the hard band emission at these temperatures.
On the other hand, the soft band contains a larger fraction of
continuum emission and is less affected.

This is just the trend seen for the hard and soft X-ray luminosities in Group 2
and 3 galaxies as $L_X^h/L_B$
decreases.
Furthermore, the change in $\log(L_X^h/L_B$) in going from
50\% solar to 10\% solar abundance is 0.6 dex, nearly the same range over which
$L_X^h/L_B$ is seen to decrease for Group 2 and 3 galaxies.
Thus, a variation in abundance for temperatures in the $0.4-0.8$ keV range
provides a simple, plausible explanation for why $L_X^s/L_B$ decreases much
more slowly than $L_X^h/L_B$.

\begin{figure}[htbp]
\vskip6.30truein
\hskip0.3truein
\includegraphics{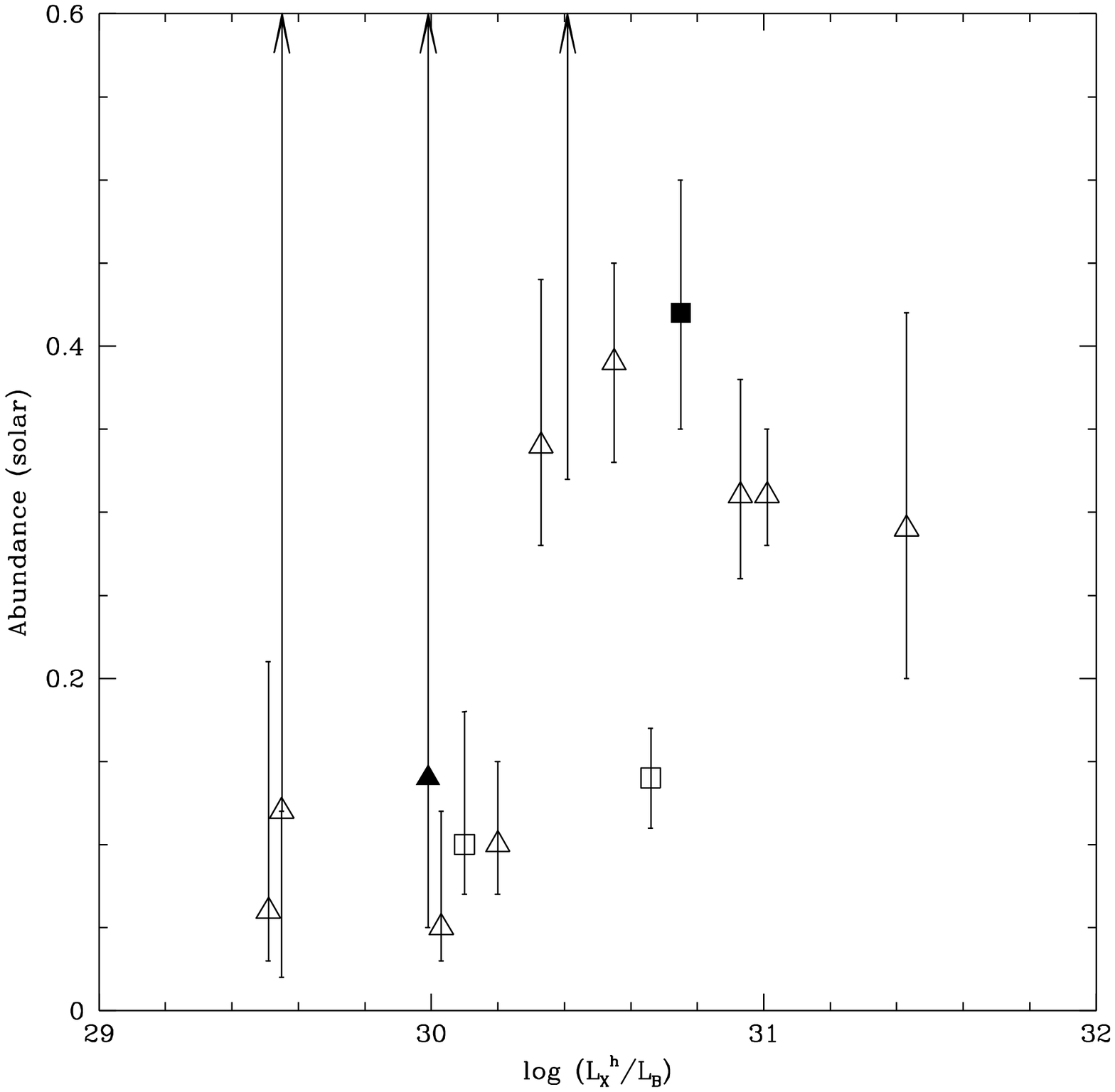}
\caption[ASCA Abundances]{
Abundances for galaxies in our sample as determined by {\it ASCA} as a function
of $\log(L_X^h/L_B$) with 90\% error bars.
Open triangles denote values from Matsumoto et al.\ (1997),
open squares denote values from Loewenstein (1996),
filled triangles denote values from Buote \& Canizares (1997),
and filled squares denote values from Arimoto et al.\ (1997).
\label{fig:asca_abund}}
\end{figure}

We have searched the literature for all elliptical and S0 galaxies in our
survey which have been observed by {\it ASCA} for which abundances have been
determined.
The abundances for these galaxies are plotted in Figure~\ref{fig:asca_abund}
versus $\log(L_X^h/L_B$).
Unfortunately, only five Group
2 and 3 galaxies have published {\it ASCA} abundances, and the 90\% errors
are rather large, so we can draw no firm conclusion on the dependence of
abundance on $L_X^h/L_B$ for galaxies in these two groups until more Group
2 and 3 spectra have been analyzed.
This Figure does suggest that the abundance does increase with
$L_X^h/L_B$ for galaxies with intermediate X-ray luminosities,
and that the abundances for Group 4 galaxies are roughly constant,
with the exception of NGC~1404. Group 4 galaxies do not exhibit the same
flattening seen in Group 2 and 3 galaxies, apparently due to the fact that
Group 4 galaxies all have nearly the same abundance; they have higher
$L_X^h/L_B$ values than the Group 2 and 3 galaxies primarily because they sit at
the center of groups of galaxies, and not because of an abundance effect.
The Group 4 galaxy NGC~1404 has an anomalously low abundance,
which would have the effect of making it underluminous in $L_X^h/L_B$
for its $L_X^s/L_B$, as is observed in Figure~\ref{fig:abund}.

In constructing the model curves in Figure~\ref{fig:abund},
one important assumption we have made is that the volume emission measure
of the hot gas is proportional to the stellar luminosity for
the Group 2 and 3 galaxies.
This would not be expected to be the case if the galaxies covered a large
range in optical luminosity, because the radii and other properties of the
galaxies would vary.
However, the average blue luminosities for Group 2 and Group 3 galaxies are
similar, $9.5\pm5.3 \times 10^{10} L_{B,\odot}$ and
$1.2\pm0.5 \times 10^{11} L_{B,\odot}$, respectively.
Furthermore,
$L_X^h \propto L_B^{1.16 \pm 0.18}$ and $L_X^s \propto L_B^{0.86 \pm 0.19}$
for Groups 2 and 3 combined.
The nearly linear dependence of X-ray luminosity with optical luminosity is
consistent with the assumption that the gas emission measure per unit stellar
luminosity mass is constant for these galaxies.
This is manifestly not true of the Group 4 galaxies, although this
may be due in part to group gas.

Finally, there are five Group 2 and 3 galaxies with significantly higher soft
X-ray luminosities than is predicted by this model. Two of these galaxies may 
really be Group 4 which have been ``misplaced" into Group 3 because of an
abnormally low abundance. As was seen with NGC~1404, a low abundance has
the effect of shifting a galaxy to the left of the best-fit line in
Figure~\ref{fig:asca_abund}. These two galaxies might have been shifted out
of the Group 4 domain because of their low abundance. It is unlikely that
abundance effects can be responsible for the three anomalous Group 2 galaxies,
though. In these galaxies, the origin of the excess soft X-ray emission is
unknown.

\subsection{Group 1 Galaxies} \label{sec:origin_group1}

As mentioned in \S~\ref{sec:colors} above, Group 1 galaxies
have very soft C21 colors, unlike their X-ray bright counterparts.
The mean color values and standard
deviations are $\langle {\rm C21} \rangle = 0.16 \pm 0.06$ and
$\langle {\rm C32} \rangle = 0.99 \pm 0.41$
(Table~\ref{tab:group_prop}).
We will use these values to test possible X-ray emission mechanisms
in these galaxies.

 From Figure~\ref{fig:c21c32em}, we see that the average colors of Group 1
galaxies could result from emission from diffuse gas with a temperature of
$\sim$0.6 keV and a metallicity below 10\% of solar.
However, there are at least three problems with such a single component
thermal model.
First, it seems unlikely that gas with such a low metallicity could exist
in an elliptical galaxy.
Assuming typical stellar mass loss rates, Fabbiano et al.\ (1994) find that
$10^8$ M$_{\odot}$ of primordial, metal-free gas would be polluted to
10\% metal enrichment by stellar winds in about
$10^7$ yrs.
Second, Fabbiano et al.\ (1994) found that, although a single component,
low abundance model could not be ruled out as a fit to the {\it ROSAT}
spectrum of the X-ray faint galaxy NGC~4382, the {\it ASCA} observation of this
galaxy was not consistent with this model.
Third, C21-C32 plot shows a few Group 1 galaxies which have colors that
are even softer than those predicted by a thermal model with zero metallicity.
Finally, if the abundance-temperature tracks shown in Figure~\ref{fig:abund}
are extended to zero metallicity, the tracks are consistent with the
luminosities of only the brightest of the Group 1 galaxies
($\log (L_X^h/L_B) \ga 29.3$).
Given the above arguments,
we will abandon the single thermal component, low abundance hypothesis.

Previous authors
(Fabbiano et al.\ 1994;
Pellegrini 1994;
Kim et al.\ 1996)
have found that a two component model also provides an adequate
fit to the spectra of low $L_X/L_B$ galaxies.
In general, the two components which are required are a soft component,
with $kT \sim 0.2$ keV and solar abundance, and a hard component with a
temperature of $\sim$5 keV. The $\sim$5 keV model component is believed to
be the integrated emission from a population of low mass X-ray binaries
(LMXBs), as is found in the bulges of spiral galaxies
(Trinchieri \& Fabbiano 1985).
Matsumoto et al.\ (1997) detected the presence of such a hard component
in many elliptical and S0 galaxies of varying $L_X/L_B$.
They also found that the X-ray luminosity of this component scales linearly
with the optical luminosity, which is consistent with a stellar origin.
The X-ray colors predicted for the hard component are
C21$ = 0.30$ and C32$ = 1.72$, if it is modeled as thermal bremsstrahlung at
a temperature of 5 keV.

The origin of the soft X-ray component is less clear.
It has been suggested that the soft component is from the integrated emission
of M star coronae,
from RS CVn binary systems,
from supersoft sources like those observed in our Galaxy, M31, and the
Magellanic clouds,
or from diffuse interstellar gas at a temperature of $0.2-0.4$ keV
(Pellegrini \& Fabbiano 1994).

The faintest elliptical galaxies are believed to have overall properties
and stellar populations similar to the bulges of large spiral galaxies.
The nearest bulges (e.g., M31) are close enough that individual X-ray
sources can be resolved.
Thus, it is useful to determine the X-ray colors and X-ray--to--optical
luminosity ratios of nearby spiral bulges;
this can provide a strong test of stellar models for the origin of
the X-ray emission in faint elliptical galaxies.

M31 is the best candidate to study because of its proximity.
To determine if M31 is similar to the Group 1 early-type galaxies in
its X-ray properties, we extracted the archival {\it ROSAT} PSPC
data on images covering the center of the bulge of this galaxy.
We determined the integrated X-ray fluxes of the bulge in our bands.
In doing so, we did not exclude point sources within the bulge,
since these sources cannot be resolved in more distant early-type
galaxies.
We have determined the $L_X^s/L_B$ and $L_X^h/L_B$ values for the inner
$10^{\prime}$ of M31, which corresponds to one effective radius
for the bulge (2 kpc at a distance of 690 kpc).
We assumed an integrated blue magnitude of 5.39 within one effective radius
(Walterbos \& Kennicutt 1988).
The total soft and hard
luminosities within 1 $r_{eff}$ assuming our standard spectral model used for
the sample were $9.54 \times 10^{38}$ ergs s$^{-1}$ and
$8.47 \times 10^{38}$ ergs s$^{-1}$, respectively. This corresponds to
$L_X^s/L_B = 29.26$ and $L_X^h/L_B = 29.21$, placing the bulge of M31
within the observed range of values for Group 1 galaxies
(Figure~\ref{fig:hard_soft}).

We also determined the X-ray colors for the bulge of M31.
For comparison to the point source catalog of Supper et al.\ (1997),
we derived the colors for the inner $5^{\prime}$ of the bulge.
We obtained (C21, C32$) = (0.13, 1.16)$, corrected for
foreground absorption ($N_H = 6.73 \times 10^{20}$ cm$^{-2}$); although
the absorption is above our cut-off limit, this should not introduce
substantial errors into the determination of the colors.
Thus, the integrated X-ray colors for the inner bulge of M31 are
consistent with the colors of the faintest X-ray galaxies
(Figure~\ref{fig:c21c32}).
In particular, the bulge of M31 has essentially the same very soft
C21 color as in elliptical and S0 galaxies.

However, the bulk of the X-ray emission from the bulge of M31 has
been resolved into a relatively small number of luminous discrete X-ray
sources.
For example, Supper et al.\ (1997) detected
22 point sources within $5^{\prime}$ of the center of M31.
They also find that these 22 sources comprise 75\% of the
total flux within $5^{\prime}$ of the center of M31. Thus, any emission from
unresolved sources or a diffuse component is not a major contributor to the
total X-ray emission within the bulge.
Supper et al.\ (1997) tabulated the count rates in the same three X-ray bands
we use to determine X-ray colors.
Three of the sources were detected only in the soft band, and are most likely
supersoft sources.
The rest of the sources are most likely LMXBs.
By summing
the count rates in the three bands for the 22 sources, we derived the X-ray
colors for the sum of emission from the discrete sources, and corrected
them for foreground absorption.
The colors obtained were (C21, C32$) = (0.08, 1.15)$.
Thus, the colors from the sum of discrete sources in the bulge of M31 are
consistent both with the integrated X-ray colors of the bulge and with
the colors of the faintest X-ray galaxies (Figure~\ref{fig:c21c32}).
Since the discrete sources provide most of the X-ray flux from
the bulge of M31, the first result is not terribly surprising.
Apparently, the source of the remaining 25\%
of unresolved flux within the bulge does not affect the colors to a great
extent.
It might raise the C21 value somewhat; if so, the diffuse emission is
harder than the discrete emission at low energies.
Removing the three supersoft sources found in the bulge does
not change the integrated X-ray colors appreciably.

The comparison between the X-ray colors of early-type galaxies and
those of the discrete sources in M31 shows that the very soft component
in ellipticals and S0s can be explained simply as emission from
luminous, discrete binary X-ray sources, which are probably mainly LMXBs.
Apparently, this was not appreciated earlier because most previous studies
assumed too simple a model for the average X-ray spectra of LMXBs
in early-type galaxies.
In particular, the colors obtained for the bulge sources of M31
(0.08, 1.15) are clearly at odds with the colors derived from a
5 keV bremsstrahlung model (0.30, 1.72), which has been commonly used to
describe the hard emission from LMXBs.

This suggests that the X-ray spectra of LMXBs contain a soft component
in addition to a $\sim$5 keV component.
As a further test of this hypothesis, we have done a more detailed analysis
of the integrated X-ray spectrum of the bulge of M31.
 From the archival data of M31, we extracted the full resolution spectrum of
the inner $5^{\prime}$ of the bulge.
The energy channels were rebinned to contain at least 20 counts,
and a background spectrum, extracted from an annulus of
$30^{\prime}-40^{\prime}$ and corrected for vignetting, was scaled to and
subtracted from the source spectrum.
The background-subtracted spectrum contained 3003 counts.
The results of the modeling the X-ray spectrum of the bulge of M31 are
summarized in Table~\ref{tab:m31}.
\begin{table}[tbp]
\caption[Spectral Fits]{}
\begin{center}
\begin{tabular}{lccccccc}
\multicolumn{8}{c}{\sc Spectral Fits to the Bulge of M31} \cr
\hline \hline
& $N_H$ & $kT_1$ & $Z_1$ & $kT_2$ & $Z_2$ & $\chi^2$ & DOF \\
& (cm$^{-2}$) & (keV) & ($Z/Z_{\odot}$) & (keV) & ($Z/Z_{\odot}$) && \\
\hline
Model 1 & $6.73 \times 10^{20}$ & 0.78$^{+0.07}_{-0.06}$ & 0$^{+0.004}_{-0.00}$
& \ldots & \ldots & 123.0 & 100 \\
Model 2 & $6.73 \times 10^{20}$ & $0.19^{+0.01}_{-0.01}$ & 1.0 (fixed) &
3.75$^{+2.16}_{-1.13}$ & 1.0 (fixed) & 163.7 & 99 \\
Model 3 & $6.73 \times 10^{20}$ & $0.36^{+0.09}_{-0.06}$ &
0.012$^{+0.012}_{-0.005}$ & $>6.4$ & 1.0 (fixed) & 100.5 & 98  \\
\hline
\end{tabular}
\end{center}
All errors are 90\% confidence limits for 1 interesting
parameter.
\label{tab:m31}
\end{table}

Previous spectral modeling of the bulge of M31 with {\it Einstein} data
by Fabbiano, Trinchieri, \& Van Speybroeck (1987) found that
the spectrum from the inner $5^{\prime}$
of the bulge was fit well by a bremsstrahlung model with $kT=13.5$ keV
and an absorbing column density somewhat less than the Galactic value.
Here, we fit the spectrum of the bulge of M31 with Raymond-Smith models.
Since the bulk of the emission from M31 is due to discrete LMXBs,
where the emission is not mainly from collisionally ionized, diffuse,
optically thin plasma, we do not attach any real physical significance
to the parameters of these fits.
Rather, we use this model to better compare to the results of previous
spectral fits to early-type galaxies, which employ the Raymond-Smith
model
(e.g., Fabbiano et al.\ 1994).

We first attempted to fit the spectrum with a one component Raymond-Smith
model with variable abundance and Galactic absorption. This yielded an
adequate fit to the data ($\chi^2=123.0/100$ degrees of freedom). The best-fit
temperature was 0.78 keV and the best-fit abundance was zero. This model is
very reminiscent of the model which adequately fit the {\it ROSAT} spectra
of several X-ray faint galaxies, but not the {\it ASCA} spectra of NGC~4382
(Fabbiano et al.\ 1994; Kim et al.\ 1996).
However, as was the case for the X-ray faint galaxies, this is not a
plausible model for the integrated emission from LMXBs, because it lacks
the very hard component seen in these objects individually and in the
{\it ASCA} spectra of elliptical galaxies.
The temperature derived for this model is significantly less than the
value determined by {\it Einstein} using a bremsstrahlung model, although
an absorbing column density less than the Galactic value was found to
yield the best fit to the {\it Einstein} data (Fabbiano et al.\ 1987),
whereas we have fixed the column density at the Galactic value.

Next, a two component Raymond-Smith model with abundances fixed at solar
was used to fit the spectra. The fit was unacceptable, with
$\chi^2=163.7$ for 99 degrees of freedom. The best-fit parameters were
$kT_1=0.19$ keV and $kT_2=3.75$ keV. This is remarkably similar to the
values obtained from the spectra of X-ray faint galaxies. However, unlike
the X-ray faint galaxies, the observation of M31 contained enough counts
to rule out this model.

We find that a vast improvement is made in the fit if we allow the abundance
of the soft component to vary. Doing so reduces $\chi^2$ to 100.5 for 98
degrees of freedom. The soft component temperature was found to be
$kT_1=0.36$ keV ($0.30-0.45$ keV) with an abundance of $<3\%$ of solar,
and the hard component temperature was $kT_2 >6.4$ keV (all errors are
quoted at the 90\% confidence level).

Due to the very low abundance of the soft component, we hesitate to attach
a physical significance to this two component model for the X-ray emission
from LMXBs.
Nonetheless, the observations do show that a soft component
should be expected from the integrated emission from LMXBs in an old
stellar population.
One does not need to invoke a separate class of X-ray--emitting objects to
account for the soft X-ray emission in X-ray faint ellipticals, especially
for the faintest of the Group 1 galaxies whose X-ray colors and $L_X/L_B$
ratios are fully consistent with those of the bulge of M31.
A definitive confirmation of this prediction will be possible after the
launch of {\it AXAF}. The backside-illuminated CCD chips of the ACIS
instrument on-board {\it AXAF} will provide the needed soft energy
response and spatial resolution to resolve LMXB X-ray emission for
early-type galaxies at a distance of Virgo or closer. 
The brightest Group 1 galaxies may contain an
additional soft component, which we will discuss in
\S~\ref{sec:origin_group1_ism}.

We have also calculated the X-ray colors of the bulge of the bright Sa 
galaxy NGC~1291 to verify the results found for M31. We have assumed
an integrated blue magnitude of 10.73 within an effective radius
of $48^{\prime\prime}$ (3.2 kpc at a distance of 13.8 Mpc;
de Vaucouleurs 1975) for this galaxy. After correcting for absorption
($2.24 \times 10^{20}$ cm$^{-2}$), we derive integrated colors of
(C21, C32$)=(0.12, 1.15)$. These values are virtually identical to those
obtained for M31. We also found that $L_X^s/L_B=29.68$ and $L_X^h/L_B=29.43$
for the bulge of NGC~1291. These values are somewhat higher than those
which were found for M31, but still within the range of observed values for
X-ray faint galaxies. A spectral fit of the bulge of NGC~1291 was performed by
Bregman, Hogg, \& Roberts (1995). They found that a two component Raymond-Smith 
model with abundances fixed at solar provided a good fit to the data, and
yielded temperature values of $kT_1=0.18$ keV and $kT_2=1.91$ keV. These values
are consistent with the temperatures derived for the bulge of M31 using
an identical spectral model (see Model 2 in Table~\ref{tab:m31}).
Although this model did not provide an adequate fit to the spectra of M31,
the observation of NGC~1291 contained only one-third the number of counts
of the M31 observation, so it is not surprising that solar
abundances could not be be ruled out for the soft component of NGC~1291.

Obviously, such a soft component in the X-ray emission from LMXBs should
be readily observable in Galactic LMXBs, if it were not for the fact that
nearly all the LMXBs in our Galaxy lie in directions of high hydrogen
column densities, because they are concentrated to the central regions of
the Galactic bulge.
Such a soft component would
be completely absorbed in these cases. However, the LMXB Her X-1 is
relatively nearby, and in a direction of low Galactic column density
($1.73 \times 10^{20}$ cm$^{-2}$). Observations of Her X-1 with
{\it ROSAT} (Mavromatakis 1993) and {\it ASCA} (Vrtilek et al.\ 1994;
Choi et al.\ 1996) find that a soft component described by a
blackbody model with $kT=0.1$ keV or a bremsstrahlung model with
$kT=0.2$ keV is needed in addition to a power law model (for the harder
X-ray emission) in order to fit the spectrum.
The X-ray colors
derived from the archived {\it ROSAT} data were found to be
(C21, C32$) = (0.08, 1.56)$ for Her X-1.
Thus, the soft component seen in X-ray faint galaxies and the bulges of
M31 and NGC~1291 also appears to be present in the spectrum of at least
one LMXB in our Galaxy, as indicated by its low C21 value.
Although the C32 value of Her X-1 is somewhat higher than that of the
Group 1 galaxies or the bulge of M31, it falls within the range of
observed C32 values found for individual LMXBs in the bulge of M31.

If NGC~5102 is excluded, the best-fit relation between $L_X^s/L_B$ and
$L_X^h/L_B$ is nearly linear (Figure~\ref{fig:hard_soft}). Along with the
fact that the X-ray colors of most of the Group 1 galaxies are consistent
with emission from stellar emission as shown above, this suggests that most
Group 1 galaxies share the same single X-ray emission mechanism, with some
galaxies having more X-ray emission per unit blue luminosity than others.
At first this may seem at odds with the results of Matsumoto et al.\ (1997),
who analyzed {\it ASCA} spectra from a sample of elliptical and S0 galaxies,
and found that all of their galaxies but one had a very hard X-ray spectral
component.
This hard component was consistent with a temperature of about
5 keV, and was attributed to the emission from LMXBs.
The luminosity of this
hard component scaled linearly with the blue luminosity of the galaxy.
We took the ratio of the $0.5-4.5$ keV X-ray luminosity per unit optical
luminosity from their paper, and converted the X-ray luminosity to
our hard band ($0.52-2.02$ keV),
assuming Model 3 of the fit for the bulge of M31 (see Table~\ref{tab:m31}).
This model gave nearly the same conversion factor as a 5 keV thermal
bremsstrahlung model, since the soft component of Model 3 contributes
very little emission in the $2.0-4.5$ keV range.
If one excludes galaxies in their sample which may harbor active nuclei,
this leads to a {\it ROSAT} hard band X-ray--to--optical ratio of
$\log(L_X^h/L_B) \approx 29.4$.
This might be a slight underestimate because of the fact that some of the
$0.5-4.5$ keV flux from the soft component of Matsumoto et al.'s (1997)
spectral fits is also from LMXB emission and has not been included in the
calculation.
This value is consistent with the luminosity ratios for the more luminous 
of the Group 1 galaxies and for the bulges of M31 and NGC~1291.

However, Figure~\ref{fig:hard_soft} shows that there are five galaxies
which have upper limits on $L_X^h/L_B$ which are at least a factor of 3
less than the value derived from the Matsumoto et al.\ (1997) correlation.
For example, NGC~5102 has an order of magnitude less hard X-ray emission than
expected for its blue luminosity, if the hard component scales linearly
with optical luminosity. The same is true for the nearby dwarf elliptical
NGC~205, which was not included in our sample because it did not meet the
requirements stated in \S~\ref{sec:galsamp}, but nonetheless had an upper limit
on log($L_X^h/L_B$) of 28.27, a factor of 10 less than expected.

One possible explanation for the discrepancy may be that the galaxies
in our sample with upper limits on the $L_X^h/L_B$ ratios significantly
less than 29.4 have fewer LMXBs per unit blue luminosity than the rest
of the galaxies in the sample. Since LMXBs contain accreting neutron stars,
the number of LMXBs in a given galaxy would decrease if fewer 
massive stars have been formed in the galaxy.
This would imply that fainter elliptical galaxies had a
steeper initial mass function.
In fact, this theory has been proposed to explain the trend of increasing
Mg/Fe ratios in elliptical galaxies as a function of increasing mass
(Worthey, Faber, \& Gonzalez 1992).
If Mg is produced primarily by Type II supernovae and Fe by Type Ia
supernovae, higher Mg/Fe ratios for larger ellipticals could imply a
higher rate of Type II supernovae and of neutron star production.
Models by Matteucci (1994) show that flattening the initial mass function
for more more massive galaxies also predicts the $M/L_B \propto L_B^{0.2}$
relation which is seen in ellipticals.

The five galaxies in the sample (plus NGC~205) with unusually low
$L_X^h/L_B$ values all have optical luminosities below
$1.7 \times 10^{10} \, L_{B,\odot}$ and are considerably fainter
(and presumably less massive) than the rest of the galaxies in the sample
(see Figure~\ref{fig:lxhlb}).
On the other hand, the Matsumoto sample consisted mostly of luminous
ellipticals.
According to the above argument,
these galaxies should have fewer high mass stars per unit blue luminosity
available to make the neutron stars present in LMXBs.
Matteucci (1994) argued that the slope of the initial mass function flattened
from $x=1.35$ (the Salpeter IMF) in fainter ellipticals to $x=0.95$ in
brighter ellipticals.
This change will produce about 2.5 times fewer stars in the mass range
$8-100 \, M_\odot$ per optical luminosity in fainter ellipticals as
compared to brighter ones.
We assume that only stars which are initially more massive than
$8 \, M_\odot$ evolve into neutron stars or black holes.
This difference might be large enough to account for the lower proportion
of LMXBs in most of the fainter Group 1 galaxies, but not in NGC~5102 or
NGC~205. It should be noted, though, that the optical luminosity of the bulge of
M31 is rather low ($\sim$$10^{10} \, L_{B,\odot}$), yet it has only a slightly
lower than expected log($L_X^h/L_B$) value (29.21).

It is also possible that fluctuations in the small number of active LMXBs
in faint early-type galaxies contribute to the variation in the
X-ray--to--optical luminosity ratio.
In particular, both NGC~5102 and NGC~205 have very low optical luminosities
($4.4 \times 10^9 L_{B,\odot}$
for NGC~5102, and $2.0 \times 10^8 L_{B,\odot}$ for NGC~205).
If the average X-ray--to--optical luminosity ratio of faint ellipticals
is $\log (L_X/L_B) \approx 29.4$, then we would expect an X-ray luminosity
of $L_X \approx 9 \times 10^{38}$ ergs s$^{-1}$ for NGC~5102 and
$L_X \approx 4 \times 10^{37}$ ergs s$^{-1}$ for NGC~205.
Particularly in the latter case, this would correspond to a small
number of LMXBs.

Next we address the issue of why Group 1 galaxies do not appear to have
a significant interstellar medium component like their X-ray bright
counterparts.
Detailed hydrodynamical simulations of the evolutionary
behavior of the gas flow in elliptical galaxies by
Loewenstein \& Mathews (1987) show that bright, massive
galaxies develop hot, gaseous halos after the Type II supernovae
rate declines early in the galaxy's history.
On the other hand, less massive galaxies are
able to sustain a wind due to Type Ia supernovae.
This wind drives the gas lost from stars out of the galaxy.
The efficiency of wind mass loss
depended on the assumed Type Ia supernovae rate and the presence and
distribution of a dark matter halo.
Later work by David, Forman, \& Jones (1991)
extended the simulations to include galaxies with luminosities as low
as $10^9 \, L_{B,\odot}$, and found basically the same results.
The hydrodynamical models of Ciotti et al.\ (1991) suggested that hot gas
in elliptical galaxies undergo a three stage (wind, outflow, inflow) evolution.
All three studies
concluded that galaxies covering the range of observed $L_X/L_B$ values
can be described in terms of differing hydrodynamical states of the gas
ranging from supersonic winds to partial winds to cooling flows, depending
on the galaxy's mass distribution and supernovae rate.
Since the Group 1 galaxies have the lowest optical luminosities and
velocity dispersions in the sample, they should have shallower potential
wells and could still be in the wind stage of their evolution.

\begin{figure}[htbp]
\vskip6.30truein
\hskip0.3truein
\includegraphics{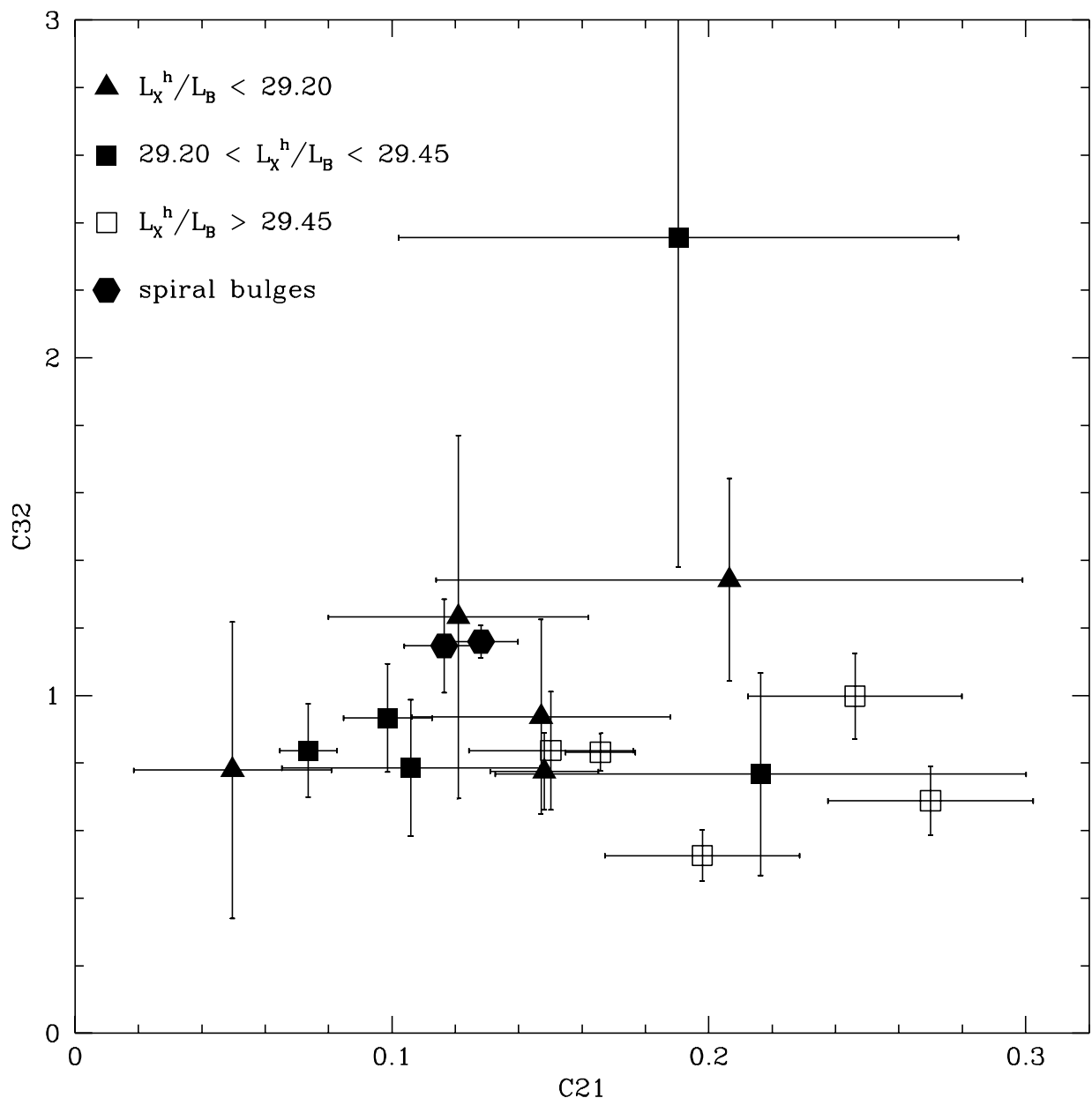}
\caption[Group 1 Colors]{
X-ray colors C21 and C32 are plotted against one another for the
Group 1 galaxies.
These galaxies have been subdivided into three $L_X^h/L_B$ classes;
the symbols and corresponding luminosity ratio ranges are shown at
the upper left.
We have also plotted the colors of the two spiral bulges
NGC~1291 (left) and M31 (right) as filled hexagons.
The faintest X-ray galaxies have colors consistent with those of the two
bulges (hexagons), but the five brightest Group 1 galaxies (open squares) have
distinctly different colors.
These five galaxies are NGC~5866, NGC~4382,
NGC~4697, NGC~4365, and NGC~1380. \label{fig:group1}}
\end{figure}

If this argument is correct, then there should exist galaxies which have
some, but not all of their ISM removed.
These galaxies would have comparable
amounts of X-ray emission from the stellar and gaseous components. If so,
these galaxies might have colors which differ somewhat from those of the
purely stellar sources, such as the bulges of M31 and NGC~1291. To search
for this we have subdivided the 15 Group 1 galaxies for which colors were
determined into three groups based on $L_X^h/L_B$.
The X-ray colors of these galaxies are shown in Figure~\ref{fig:group1}.
The two lowest $L_X^h/L_B$ subgroups among Group 1
galaxies (filled triangles and filled squares) have colors which scatter
around those of the spiral bulges (hexagons), at least within the errors.
On the other hand,
the highest $L_X^h/L_B$ subgroup (open squares) have colors which are
distinctly different than those of the bulges.
This could be explained by the presence of an additional
soft component with C$21 \ga 0.25$ and C$32 \la 0.7$.
This extra soft component
would naturally explain why these galaxies have higher $L_X/L_B$ values than
the X-ray faintest galaxies. Below, we show that an
ISM provides the most reasonable explanation for the additional soft
component.

\section{Alternatives for the Soft Component in Group 1 Galaxies}
\label{sec:soft_alternatives}

Previous authors (Kim et al.\ 1992; Pellegrini \& Fabbiano 1994) have
suggested that the integrated emission from M star coronae, RS CVn
binary stars, supersoft sources, or a warm ISM is responsible for the
soft component found in Group 1 galaxies, since these sources have
the soft X-ray properties needed to explain the soft emission. However,
the general consensus was that none of these mechanisms is entirely
responsible for the soft X-ray emission. Earlier, we showed that LMXBs
are probably responsible for the soft emission in the X-ray faintest galaxies
and not the sources mentioned above. Here, we present further arguments as
to why these sources cannot be substantial contributors to the X-ray emission
in the X-ray faintest galaxies, although a warm ISM may be present in
the brighter Group 1 galaxies.

\subsection{M Star Coronae} \label{sec:origin_group1_Mstar}

Although intrinsically faint X-ray sources, M dwarf stars can be a source
of substantial X-ray emission because of the sheer number of M stars relative
to dwarf stars of earlier spectral types. M stars are also known to possess
soft X-ray characteristics (Giampapa et al.\ 1996).
However, it is unlikely that M dwarf stars are bright enough to produce the
required X-ray luminosity. As pointed out by Pellegrini \& Fabbiano (1994),
if one assumes a Salpeter initial mass function and an appropriate
mass-to-light ratio for elliptical galaxies, the average X-ray luminosity
per M star must be $2 \times 10^{28}$ ergs s$^{-1}$ to account for the
very soft X-ray emission in the X-ray faint galaxy NGC~4365.
Schmitt, Fleming, \& Giampapa (1995) show
that the average X-ray luminosity of stars in the solar neighborhood is
$\sim$$3 \times 10^{27}$ ergs s$^{-1}$.
This is likely to be a significant overestimate of the
average X-ray luminosity of M stars in old stellar systems, as the X-ray
luminosity of stars is expected to decrease with age as the rotation of the
star (and hence magnetic activity which produces the X-ray emission) decreases
(Skumanich 1972).

Another test of the average X-ray luminosity of older M dwarf stars is provided
by globular clusters.
Given that globular clusters and elliptical galaxies possess similar
stellar populations, X-ray emission from any stellar source should scale with
optical luminosity between the two systems.
To search for a diffuse X-ray component in globulars,
we extracted the {\it ROSAT} PSPC observation of NGC~6752 from the archive.
We selected this globular cluster because of its low Galactic column
density ($N_H = 2.40 \times 10^{20}$ cm$^{-2}$; Stark et al.\ 1992),
its proximity (distance of 4.2 kpc), and its bright
apparent blue magnitude, $m_B = 5.99$, which corresponds to a luminosity
of $1.1 \times 10^5 \, L_{B,\odot}$
(Peterson 1993).
After cleaning, the exposure was 4939 s.
The X-ray luminosity in the soft band was determined within the half-light
radius of $115^{\prime\prime}$ (Trager, Djorgovski, \& King 1993), using
the same spectral model as was used for the galaxy sample.
A $1\sigma$ upper limit of $9.48 \times 10^{31}$ ergs s$^{-1}$ was obtained
for the $0.11-0.41$ keV luminosity within the half-light radius,
corresponding to an upper limit on $L_X^s/L_B$ of 27.23.
This is a factor of nearly
300 less than the $L_X^s/L_B$ value of the X-ray faint galaxies NGC~4365
and NGC~4382.

These limits on the diffuse X-ray emission from globular
clusters also apply to any other stellar source for the soft X-ray
emission in Group 1 ellipticals.
There are, however, at least two ways in which this limit might be
circumvented.
First, the central regions of elliptical galaxies have very high stellar
metallicities,
much higher than even the most metal rich globular clusters
(Aaronson et al.\ 1978).
Thus, if there were a stellar component whose X-ray--to--optical luminosity
ratio increased extremely rapidly with the metallicity of the stars,
then the globular cluster limits would not apply.
Since elliptical galaxies have metallicities which decrease with
galactocentric radius
(Schombert et al.\ 1993),
this model would predict that the X-ray emission in the faintest
elliptical galaxies is more centrally condensed than the optical emission.
For the X-ray faint galaxy NGC~4382, the emission in both X-ray bands
seems to be somewhat more extended than the optical light.
This model would also require that the X-ray colors of Group 1 galaxies
become harder with increasing radius.
For the few Group 1 galaxies with enough counts to derive radial color
profiles, this does not seem to be the case (see \S~\ref{sec:radcolor}).

Another possible exception to the limits from globular clusters is that
they do not apply to any stellar component which is the result of rare but
very luminous X-ray sources.
Such sources might be too uncommon to have been found in the globular
clusters searched.
Given the typical blue optical luminosities of the Group 1 galaxies of
$5.8 \times 10^{10} \, L_{B,\odot}$
and the blue luminosities of the globular clusters of
$10^5 L_{B,\odot}$,
such rare sources would have to be more luminous in X-rays than
$L_X \ga 10^{35}$ ergs s$^{-1}$.
If the X-ray luminosity were less, the
number of such objects required in galaxies would imply that
one such object would have been observed in most globular clusters
observed to date.
In fact, one possible candidate (a supersoft source in M3) has been
found among all of the globulars observed
(Verbunt et al.\ 1994; \S~\ref{sec:origin_group1_sss}).

\subsection{RS CVn Systems} \label{sec:origin_group1_RSCVn}

RS CVn systems are composed of a G or K giant or
subgiant and a late-type main sequence or subgiant companion.
A similar type of binary system are BY Draconis stars, composed of two
late-type main sequence stars.
In principle, these systems have some advantages as X-ray sources in
elliptical galaxies when compared to single stars. The rapid rotation
brought on by the synchronization of the rotation period to the
orbital period leads to substantially higher magnetic activity (and
hence X-ray emission) in these systems than in single old main sequence stars.

Dempsey et al.\ (1993) find that RS CVns have X-ray luminosities which range
from $10^{29}-10^{31.5}$ ergs s$^{-1}$ and spectra which are described by a
two-temperature Raymond Smith model: a low temperature (0.2 keV) component
and a high temperature (1.5 keV) component, with the high temperature
component having a volume emission measure about 2.5 times that of the
low temperature component.
Although the presence of a $\sim$0.2 keV component appears promising in
explaining the soft component in ellipticals and S0s, the two temperature
model predicts colors of (0.26, 1.51).
The C32 color is significantly harder than that observed for Group 1 galaxies.
Also, the C21 colors of RS CVns are also somewhat harder than those observed
for the faintest Group 1 galaxies, and the addition of a third component
needed to lower the C32 value will raise C21 even more.
Thus, it appears unlikely that RS CVn systems can contribute appreciably
to the X-ray emission of X-ray faint early-type galaxies.
BY Draconis binary systems
have nearly identical spectral properties to
RS CVn stars (Dempsey et al.\ 1997), and are also unlikely
candidates for the source of the soft X-ray emission in Group 1 galaxies.

\subsection{Supersoft Sources} \label{sec:origin_group1_sss}

A relatively new class of X-ray emitting objects first observed with
{\it Einstein} and later confirmed with {\it ROSAT} are supersoft sources.
Supersoft sources have been detected in our Galaxy
(\"Ogelman et al.\ 1993),
M31 (Supper et al.\ 1997),
and the Magellanic clouds (Kahabka, Pietsch, \& Hasinger 1994).
Little is known about the distribution and relative number of these sources
among Hubble types, so it it possible that early-type galaxies can contain
a significantly larger number of them than spiral galaxies.
Supersoft sources are characterized by a lack of emission above $\sim$0.4
keV, and are believed to be the result of a white dwarf burning matter
accreted from a companion star (van den Heuvel et al.\ 1992).
Supersoft source spectra can be fit by a blackbody model with a temperature
of $10-50$ eV.
They typically have luminosities in the {\it ROSAT} band of
a few $ \times 10^{36}$ erg s$^{-1}$ (Kahabka et al.\ 1994).
Thus, $\sim$$10^4$ such sources would be necessary to account for the soft
X-ray emission in the most X-ray luminous Group 1 galaxies.
These sources are bright enough and rare enough that we would not
expect them to occur commonly in globular clusters;
thus, they would not be subject to the limits on the diffuse soft
emission from globulars in \S~\ref{sec:origin_group1_Mstar}.
For a typical globular cluster with an optical luminosity of
$L_B \approx 10^5 \, L_\odot$, an X-ray--to--optical luminosity ratio
of $\log ( L_X^h / L_B ) = 29.7$ implies an X-ray luminosity of
$L_X^s \approx 5 \times 10^{34}$ ergs s$^{-1}$, or 1/20 of an
$L_X^s = 10^{36}$ ergs s$^{-1}$ supersoft source.
In fact, 
only one supersoft source has been observed within a globular cluster;
the globular cluster M3 was observed to harbor such a supersoft source
(Verbunt et al.\ 1994), albeit a rather faint one
($L_X\sim 10^{35}$ ergs s$^{-1}$).

However, the colors predicted by such a component are far too low to match
the observed values.
A blackbody model with a temperature of 50 eV yields
(C21, C32$)=(0.005, 0.01)$.
Thus, the addition of supersoft sources will diminish the C21 value
obtained for bulges, which is in the opposite sense of what is observed
for the brighter Group 1 galaxies.

\subsection{Warm Interstellar Gas} \label{sec:origin_group1_ism}

As mentioned above, if the X-ray faintest galaxies have lost their
interstellar medium due to winds, there should
exist some galaxies which have had some but not all of their ISM removed.
The five Group 1 galaxies with the highest $L_X^h/L_B$ values may
represent this class. Perhaps a few of the faintest Group 2 galaxies are
members of this class also. The colors of the five Group 1 galaxies
suspected to have some ISM are (C21, C32$)=(0.15-0.30, 0.50-1.0)$ and
the colors of the stellar component as derived from M31 are
(C21, C32$)=(0.08,1.15)$. Depending on the
relative contributions from the stellar and gaseous components, the colors
of the gaseous component required to yield the observed colors would be in the
range (C21, C32$)=(0.25-0.50, 0.20-0.85)$. Judging from the color
models of Figure~\ref{fig:c21c32em}, a thermal model with a temperature
of $0.3-0.6$ keV and a metallicity of $10-20\%$ of solar would be capable
of producing the required colors when combined with the emission from
stellar sources. This is the range of temperatures of metallicities that
would be expected from interstellar gas in low luminosity systems,
if one extrapolates from the brighter galaxies.

Among the three galaxies for which color profiles could be derived, there
is no consensus concerning the radial color trends. NGC~4382 shows a constant
C21 and C32 profile. NGC~4697 shows a constant C21 profile and a decreasing
C32 profile. NGC~4365 shows a decreasing C21 profile and a constant C32
profile. This lack of agreement could mean that the ISM component is
more extended than the stellar component in some galaxies and less
extended in others. Alternatively, gradients in the temperature and
abundance of the ISM component might also be responsible for the color 
gradients.

\section{Conclusions} \label{sec:conclusions}

We have analyzed a large elliptical and S0 galaxy sample observed with the
{\it ROSAT} PSPC.
We derived X-ray luminosities and X-ray--to--optical luminosity ratios in
soft and hard {\it ROSAT} bands.
We defined two X-ray colors;
integrated values and radial profiles in these two colors were determined
for the galaxies in our sample.
We find a very large range in values (a factor of $\sim$500) for the
X-ray--to--optical
luminosity ratio for galaxies in our sample, as has generally been
found in the past.
The X-ray colors, the color profiles, and the ratio of hard to soft
band luminosities all vary as a function of $L_X / L_B$, which suggests
that different emission processes are important in different galaxies.
We divide the galaxies into four groups based on their X-ray--to--optical
luminosity ratio:
Group 1, $\log(L^h_X/L_B) < 29.7$;
Group 2, $29.7 \le \log(L^h_X/L_B) < 30.0$;
Group 3, $30.0 \le \log(L^h_X/L_B) < 30.4$;
Group 4, $\log(L^h_X/L_B) \ge 30.4$.

The X-ray brightest galaxies (Group 4) have colors and luminosities which
are consistent with thermal emission from hot gas at temperatures around
0.8 keV
and abundances of approximately one-half solar.
These galaxies have hard and soft band X-ray luminosities which increase
in proportion to one another, and which increase rapidly with the
optical luminosity.
They often sit at the centers of groups, and there is evidence that
this increases their X-ray luminosities.
The galaxies in deeper potential wells may have higher X-ray
luminosities per unit blue luminosity because they have lost less gas during
their early evolution, or have accreted group gas.

The intermediate X-ray luminosity galaxies (Groups 2 and 3) have colors
which are generally consistent with thermal emission from lower temperature
and abundance gas than the Group 4 galaxies.
For these galaxies, the soft band X-ray luminosity--to--optical luminosity
ratios ($L_X^s/L_B$) are nearly constant from galaxy to galaxy, while
the hard band luminosity drops more rapidly than the optical luminosity.
One explanation for this trend would be if the hard band luminosity were
dominated by thermal emission by gas, and the soft band luminosity were
dominated by stellar sources.
This would require that the thermal X-ray luminosity decline rapidly
with decreasing optical luminosity, as is seen for the brighter
Group 1 galaxies.
However, this would not explain why the soft band X-ray--to--optical
luminosity ratio $L_X^s/L_B$ continues to decrease for
$\log(L_X^h/L_B) \la 29.7$.
Also, the {\it ROSAT} and {\it ASCA} X-ray spectra of these galaxies
suggest that the bulk of the emission is thermal emission from hot gas
with subsolar abundances
(e.g., Davis \& White 1996; Matsumoto et al.\ 1997).
Alternatively, the bulk of the X-ray luminosity might be thermal in origin,
with $L_X^h/L_B$ decreasing as the gaseous iron abundance decreases.
We show that lowering the iron abundance mainly reduces the {\it ROSAT}
hard band emission, because this emission is dominated by the iron $L$-line
complex at $\sim$1 keV.
	
The X-ray emission from most of the fainter Group 1 galaxies can best
be described as the integrated emission from low mass X-ray binaries as
was seen to be the case for the bulge of M31. In these galaxies, the ISM could
have been removed by supernovae-driven winds.
The X-ray colors and X-ray--to--optical
luminosity ratios for all but a small subset of the Group 1 galaxies
were consistent with the colors of the bulges of M31 and NGC~1291.
The soft component previously seen in these galaxies was shown to be the
result of the more complex spectrum of LMXBs;
they are not well-represented by a simple, single temperature 5 keV
bremsstrahlung model, which had been previously assumed for LMXBs.
The X-ray colors of the nearby LMXB Her X-1 confirm the existence
of a soft X-ray component in at least one Galactic LMXB.
Individual LMXBs in the faintest galaxies should be
resolvable with {\it AXAF} for galaxies as far away as Virgo.

As supporting evidence that the soft component is from LMXBs, we show that the
three most likely alternative stellar sources of the soft emission
are unlikely to be important contributors.
We find that M stars can match the X-ray colors of Group 1 ellipticals,
but that they are probably too faint.
We set a limit on the cumulative X-ray luminosity of old M stars from
upper limits on the diffuse X-ray emission from globular clusters.
This gives X-ray luminosities which are a factor of
at least 300 lower than the soft band X-ray luminosities of Group 1
ellipticals.
(However, this limit might not apply if the X-ray emission from
M stars increased strongly with metallicity.)
RS CVn stars have C32 colors which are too hard to match the observed
colors of ellipticals, and adding hard emission from LMXBs only increases
the discrepancy.
Supersoft sources are too soft to explain the observed colors of
ellipticals.

The brightest of the Group 1 galaxies might contain significant amounts
of interstellar matter which has not been completely removed from the galaxy.
These galaxies have higher C21 values and lower C32 values than the other
Group 1 galaxies.
This would account for their higher $L_X/L_B$ values as compared to
the faintest Group 1 galaxies and the bulges.
If the ISM has a temperature of $0.3-0.6$ keV and low metallicities, its
emission would produce the observed colors when mixed with the right
proportion of stellar emission.

Finally, we find that the faintest X-ray galaxies have hard band
{\it ROSAT} X-ray luminosities which are significantly less than expected
from the integrated emission of LMXBs,
as derived from brighter elliptical galaxies
(Matsumoto et al.\ 1997)
and spiral galaxy bulges.
This suggests that the fractional population or luminosity of LMXBs may
be lower in less luminous galaxies.
One way to produce such a variation would be if the slope of the
initial mass function in early-type galaxies became flatter as the
optical luminosity increased.
Small number statistical
variations in the number of LMXBs in these galaxies may also play a role.

\acknowledgments

We thank Michael Loewenstein and an anonymous referee for many very
useful comments and suggestions.
We also thank Steve Balbus, Tim Kallman, Bob O'Connell, Richard Mushotzky,
and Jean Swank for useful discussions.
This research has made use of data obtained through the High Energy
Astrophysics Science Archive Research Center Online Service, provided
by the NASA/Goddard Space Flight Center.
J. A. I. and C. L. S. were supported in part by NASA ROSAT grant NAG 5--3308,
and ASCA grant NAG 5-2526.
C. L. S. was also supported by NASA Astrophysical Theory Program grant 5-3057.
J. A. I. was supported by the Achievement Rewards for College Scientists
Fellowship, Metropolitan Washington Chapter.

\clearpage
\end{document}